\documentclass[12pt,a4paper,twoside]{article}
\usepackage{epsfig}

\catcode`\@=11
\def\lop#1\to#2{\expandafter\lopoff#1\lopoff#1#2}
\long\def\lopoff,#1,#2\lopoff#3#4{\def#4{#1}\def#3{,#2}}
\def\@@mlistempty{,}
\newif\iflistnonempty
\def\multiputlist(#1,#2)(#3,#4){\@ifnextchar
[{\@imultiputlist(#1,#2)(#3,#4)}{\@imultiputlist(#1,#2)(#3,#4)[]}}

\long\def\@imultiputlist(#1,#2)(#3,#4)[#5]#6{{%
\@xdim=#1\unitlength \@ydim=#2\unitlength
\listnonemptytrue \def\@@mlist{,#6,} % need this for end condition
\loop
\lop\@@mlist\to\@@firstoflist
\@killglue\raise\@ydim\hbox to\z@{\hskip
\@xdim\@imakepicbox(0,0)[#5]{\@@firstoflist}\hss}
\advance\@xdim #3\unitlength\advance\@ydim #4\unitlength
\ifx\@@mlist\@@mlistempty \listnonemptyfalse\fi
\iflistnonempty
\repeat\relax
\ignorespaces}}
\newcount\@@multicnt
\def\matrixput(#1,#2)(#3,#4)#5(#6,#7)#8#9{%
\ifnum#5>#8\@matrixput(#1,#2)(#3,#4){#5}(#6,#7){#8}{#9}%
\else\@matrixput(#1,#2)(#6,#7){#8}(#3,#4){#5}{#9}\fi}

\long\def\@matrixput(#1,#2)(#3,#4)#5(#6,#7)#8#9{{\@killglue%
\@multicnt=#5\relax\@@multicnt=#8\relax%
\@xdim=0pt%
\@ydim=0pt%
\setbox\@tempboxa\hbox{\@whilenum \@multicnt > 0\do {%
\raise\@ydim\hbox to \z@{\hskip\@xdim #9\hss}%
\advance\@multicnt \m@ne%
\advance\@xdim #3\unitlength\advance\@ydim #4\unitlength}}%
\@xdim=#1\unitlength%
\@ydim=#2\unitlength%
\@whilenum \@@multicnt > 0\do {%
\raise\@ydim\hbox to \z@{\hskip\@xdim \copy\@tempboxa\hss}%
\advance\@@multicnt \m@ne%
\advance\@xdim #6\unitlength\advance\@ydim #7\unitlength}%
\ignorespaces}}
\newcount\d@lta
\newdimen\@delta
\newdimen\@@delta
\newcount\@gridcnt
\def\grid(#1,#2)(#3,#4){\@ifnextchar [{\@igrid(#1,#2)(#3,#4)}%
{\@igrid(#1,#2)(#3,#4)[@,@]}}

\long\def\@igrid(#1,#2)(#3,#4)[#5,#6]{%
\makebox(#1,#2){%
\@delta=#1pt\@@delta=#3pt\divide\@delta \@@delta\d@lta=\@delta%
\advance\d@lta \@ne\relax\message{grid=\the\d@lta\space x}%
\multiput(0,0)(#3,0){\d@lta}{\hbox to\z@{\hskip -\@halfwidth \vrule
	 \@width \@wholewidth \@height #2\unitlength \@depth \z@\hss}}%
\ifx#5@\relax\else%
\global\@gridcnt=#5%
\multiput(0,0)(#3,0){\d@lta}{%
\makebox(0,-2)[t]{\number\@gridcnt\global\advance\@gridcnt by #3}}%
\global\@gridcnt=#5%
\multiput(0,#2)(#3,0){\d@lta}{\makebox(0,0)[b]{\number\@gridcnt\vspace{2mm}%
\global\advance\@gridcnt by #3}}%
\fi%
\@delta=#2pt\@@delta=#4pt\divide\@delta \@@delta\d@lta=\@delta%
\advance\d@lta \@ne\relax\message{\the\d@lta . }%
\multiput(0,0)(0,#4){\d@lta}{\vrule \@height \@halfwidth \@depth \@halfwidth
	 \@width #1\unitlength}%
\ifx#6@\relax\else
\global\@gridcnt=#6%
\multiput(0,0)(0,#4){\d@lta}{%
\makebox(0,0)[r]{\number\@gridcnt\ \global\advance\@gridcnt by #4}}%
\global\@gridcnt=#6%
\multiput(#1,0)(0,#4){\d@lta}{%
\makebox(0,0)[l]{\ \number\@gridcnt\global\advance\@gridcnt by #4}}%
\fi}}
\def\picsquare{\hskip -0.5\@wholewidth%
\vrule height \@halfwidth depth \@halfwidth width \@wholewidth}
\def\picsquare@bl{\vrule height \@wholewidth depth \z@  width \@wholewidth}
\newif\if@jointhem \global\@jointhemfalse
\newif\if@firstpoint \global\@firstpointtrue
\newcount\@joinkind
\def\dottedjoin{\global\@jointhemtrue \global\@joinkind=0\relax
  \bgroup\@ifnextchar[{\@idottedjoin}{\@idottedjoin[\picsquare@bl]}}
\def\@idottedjoin[#1]#2{\gdef\dotchar@join{#1}\gdef\dotgap@join{#2}}
\def\enddottedjoin{\global\@jointhemfalse \global\@firstpointtrue\egroup}
\def\dashjoin{\global\@jointhemtrue \global\@joinkind=1\relax
  \bgroup\@ifnextchar[{\@idashjoin}{\@idashjoin[\dashlinestretch]}}
\def\@idashjoin[#1]#2{\edef\dashlinestretch{#1}\gdef\dashlen@join{#2}%
\@ifnextchar[{\@iidashjoin}{\gdef\dotgap@join{}}}
\def\@iidashjoin[#1]{\gdef\dotgap@join{#1}}

\def\drawjoin{\global\@jointhemtrue \global\@joinkind=2\relax
  \bgroup\@ifnextchar[{\@idrawjoin}{}}
\def\@idrawjoin[#1]{\def\drawlinestretch{#1}}

\long\def\jput(#1,#2)#3{{\@killglue\raise#2\unitlength\hbox to \z@{\hskip
#1\unitlength #3\hss}\ignorespaces}
\if@jointhem
 \if@firstpoint \gdef\x@one{#1} \gdef\y@one{#2} \global\@firstpointfalse
 \else\ifcase\@joinkind
	\@dottedline[\dotchar@join]{\dotgap@join\unitlength}%
(\x@one\unitlength,\y@one\unitlength)(#1\unitlength,#2\unitlength)
	\or\@dashline[\dashlinestretch]{\dashlen@join}[\dotgap@join]%
(\x@one,\y@one)(#1,#2)
	\else\@drawline[\drawlinestretch](\x@one,\y@one)(#1,#2)\fi
    \gdef\x@one{#1} \gdef\y@one{#2}
 \fi
\fi}
\newdimen\@dotgap
\newdimen\@ddotgap
\newcount\@x@diff
\newcount\@y@diff
\newdimen\x@diff
\newdimen\y@diff
\newbox\@dotbox
\newcount\num@segments
\newcount\num@segmentsi
\newif\ifsqrt@done
\def\sqrtandstuff#1#2#3{
\ifdim #1 <0pt \@x@diff= -#1 \else\@x@diff=#1\fi
\ifdim #2 <0pt \@y@diff= -#2 \else\@y@diff=#2\fi
\@dotgap=#3 \divide\@dotgap \tw@
\advance\@x@diff \@dotgap \advance\@y@diff \@dotgap% for round-off errors
\@dotgap=#3
\divide\@x@diff \@dotgap \divide\@y@diff \@dotgap
\sqrt@donefalse
\ifnum\@x@diff < 2
   \ifnum\@y@diff < 2 \num@segments=\@x@diff \advance\num@segments \@y@diff
		      \sqrt@donetrue
        \else\num@segments=\@y@diff \sqrt@donetrue\fi
   \else\ifnum\@y@diff < 2 \num@segments=\@x@diff \sqrt@donetrue\fi
\fi
\ifsqrt@done \ifnum\num@segments=\z@ \num@segments=\@ne\fi\relax
 \else \ifnum\@y@diff >\@x@diff
		 \@tempcnta=\@x@diff \@x@diff=\@y@diff \@y@diff=\@tempcnta
       \fi    		%exchange @x@diff & @y@diff, so now @x@diff > @y@diff
  \num@segments=\@y@diff
  \multiply\num@segments \num@segments
  \multiply\num@segments by 457
  \divide\num@segments \@x@diff
  \advance\num@segments by 750 % for round-off, going to divide by 1000.
  \divide\num@segments \@m
  \advance\num@segments \@x@diff
\fi}
\def\dottedline{\@ifnextchar [{\@idottedline}{\@idottedline[\picsquare@bl]}}
\def\@idottedline[#1]#2(#3,#4){\@ifnextchar (%
{\@iidottedline[#1]{#2}(#3,#4)}{\relax}}
\def\@iidottedline[#1]#2(#3,#4)(#5,#6){\@dottedline[#1]{#2\unitlength}%
(#3\unitlength,#4\unitlength)(#5\unitlength,#6\unitlength)%
\@idottedline[#1]{#2}(#5,#6)}
\long\def\@dottedline[#1]#2(#3,#4)(#5,#6){{%
\x@diff=#5\relax\advance\x@diff by -#3\relax
\y@diff=#6\relax\advance\y@diff by -#4\relax
\sqrtandstuff{\x@diff}{\y@diff}{#2}
\divide\x@diff \num@segments
\divide\y@diff \num@segments
\advance\num@segments \@ne     % to put the last point at destination.
\setbox\@dotbox\hbox{#1}% just to get the dimensions of the character.
\@xdim=#3 \@ydim=#4
\ifdim\ht\@dotbox >\z@% otherwise its a circle.
  \advance\@xdim -0.5\wd\@dotbox
  \advance\@ydim -0.5\ht\@dotbox
  \advance\@ydim .5\dp\@dotbox\fi
\@killglue
\loop \ifnum\num@segments > 0
\unskip\raise\@ydim\hbox to\z@{\hskip\@xdim #1\hss}%
\advance\num@segments \m@ne\advance\@xdim\x@diff\advance\@ydim\y@diff%
\repeat
\ignorespaces}}
\def\dashlinestretch{0} %well, could have used a counter.
\def\dashline{\@ifnextchar [{\@idashline}{\@idashline[\dashlinestretch]}}
\def\@idashline[#1]#2{\@ifnextchar [{\@iidashline[#1]{#2}}%
{\@iidashline[#1]{#2}[\@empty]}} %\@empty needed-- later checked with \ifx 
\def\@iidashline[#1]#2[#3](#4,#5){\@ifnextchar (%
{\@iiidashline[#1]{#2}[#3](#4,#5)}{\relax}}
\def\@iiidashline[#1]#2[#3](#4,#5)(#6,#7){%
\@dashline[#1]{#2}[#3](#4,#5)(#6,#7)%
\@iidashline[#1]{#2}[#3](#6,#7)}
\long\def\@dashline[#1]#2[#3](#4,#5)(#6,#7){{%
\x@diff=#6\unitlength \advance\x@diff by -#4\unitlength
\y@diff=#7\unitlength \advance\y@diff by -#5\unitlength
\@tempdima=#2\unitlength \advance\@tempdima -\@wholewidth
\sqrtandstuff{\x@diff}{\y@diff}{\@tempdima}
\ifnum\num@segments <3 \num@segments=3\fi% min number of dashes I can plot
\@tempdima=\x@diff \@tempdimb=\y@diff
\divide\@tempdimb by\num@segments
\divide\@tempdima by\num@segments
{\ifx#3\@empty \relax
    \ifdim\@tempdima < 0pt \x@diff=-\@tempdima\else\x@diff=\@tempdima\fi
    \ifdim\@tempdimb < 0pt \y@diff=-\@tempdimb\else\y@diff=\@tempdimb\fi
    \ifdim\x@diff < 0.3pt %it's a vertical dashline
           \ifdim\@tempdimb > 0pt
	        \global\setbox\@dotbox\hbox{\hskip -\@halfwidth \vrule
		 \@width \@wholewidth \@height \@tempdimb}
	   \else\global\setbox\@dotbox\hbox{\hskip -\@halfwidth \vrule
		 \@width \@wholewidth \@height\z@ \@depth -\@tempdimb}\fi
       \else\ifdim\y@diff < 0.3pt %it's a horizontal dashline
               \ifdim\@tempdima >0pt
		  \global\setbox\@dotbox\hbox{\vrule \@height \@halfwidth
		 		\@depth \@halfwidth \@width \@tempdima}
		\else\global\setbox\@dotbox\hbox{\hskip \@tempdima
			 \vrule \@height \@halfwidth \@depth \@halfwidth
				 \@width -\@tempdima \hskip \@tempdima}\fi
	    \else\global\setbox\@dotbox\hbox{%
\@dottedline[\picsquare]{0.98\@wholewidth}(0pt,0pt)(\@tempdima,\@tempdimb)}
\fi\fi
\else\global\setbox\@dotbox\hbox{%
\@dottedline[\picsquare]{#3\unitlength}(0pt,0pt)(\@tempdima,\@tempdimb)}
\fi}
\advance\x@diff by -\@tempdima % both have same sign
\advance\y@diff by -\@tempdimb
\@tempdima=\num@segments\@wholewidth \@tempdima=2\@tempdima 
\@tempcnta=\@tempdima \@tempdima=#2\unitlength \@tempdimb=0.5\@tempdima
\@tempcntb=\@tempdimb \advance\@tempcnta by \@tempcntb % round-off error
\divide\@tempcnta by\@tempdima \advance\num@segments by -\@tempcnta
\ifnum #1=0 \relax\else\ifnum #1 < -100
  \typeout{***dashline: reduction > -100 percent implies blankness!***}
\else\num@segmentsi=#1 \advance\num@segmentsi by 100
     \multiply\num@segments by\num@segmentsi \divide\num@segments by 100
\fi\fi
\divide\num@segments by 2 % earlier num@segments included 'empty dashes' too.
\ifnum\num@segments >0 % if =0 then don't divide => \x@diff & \y@diff
 \divide\x@diff by\num@segments%   remain same.
 \divide\y@diff by\num@segments
 \advance\num@segments by\@ne %for the last segment for which I subtracted
 \else\num@segments=2 % one at each end.
\fi
\@xdim=#4\unitlength \@ydim=#5\unitlength
\@killglue
\loop \ifnum\num@segments > 0
\unskip\raise\@ydim\hbox to\z@{\hskip\@xdim \copy\@dotbox\hss}%
\advance\num@segments \m@ne\advance\@xdim\x@diff\advance\@ydim\y@diff%
\repeat
\ignorespaces}}
\newif\if@flippedargs
\def\lineslope(#1,#2){%
\ifdim #1 <0pt \@xdim= -#1 \else\@xdim=#1\fi
\ifdim #2 <0pt \@ydim= -#2 \else\@ydim=#2\fi
\ifdim\@xdim >\@ydim \@tempdima=\@xdim \@xdim=\@ydim \@ydim=\@tempdima
\@flippedargstrue\else\@flippedargsfalse\fi% x < y
\ifdim\@ydim >1pt \@tempcnta=\@ydim
            \divide\@tempcnta by 65536% now \@tempcnta=integral part of #1.
            \divide\@xdim \@tempcnta\fi
\ifdim\@xdim <.083333pt \@xarg=1 \@yarg=0
 \else\ifdim\@xdim <.183333pt	\@xarg=6 \@yarg=1
 \else\ifdim\@xdim <.225pt 	\@xarg=5 \@yarg=1
 \else\ifdim\@xdim <.291666pt 	\@xarg=4 \@yarg=1
 \else\ifdim\@xdim <.366666pt 	\@xarg=3 \@yarg=1
 \else\ifdim\@xdim <.45pt 	\@xarg=5 \@yarg=2
 \else\ifdim\@xdim <.55pt 	\@xarg=2 \@yarg=1
 \else\ifdim\@xdim <.633333pt 	\@xarg=5 \@yarg=3
 \else\ifdim\@xdim <.708333pt 	\@xarg=3 \@yarg=2
 \else\ifdim\@xdim <.775pt 	\@xarg=4 \@yarg=3
 \else\ifdim\@xdim <.816666pt 	\@xarg=5 \@yarg=4
 \else\ifdim\@xdim <.916666pt 	\@xarg=6 \@yarg=5
       \else			\@xarg=1 \@yarg=1%
\fi\fi\fi\fi\fi\fi\fi\fi\fi\fi\fi\fi
\if@flippedargs\relax\else\@tempcnta=\@xarg \@xarg=\@yarg
			  \@yarg=\@tempcnta\fi
\ifdim #1 <0pt \@xarg= -\@xarg\fi
\ifdim #2 <0pt \@yarg= -\@yarg\fi
}
\newif\if@toosmall
\newif\if@drawit
\newif\if@horvline
\def\drawlinestretch{0} %well, could have used a counter.
\def\drawline{\@ifnextchar [{\@idrawline}{\@idrawline[\drawlinestretch]}}
\def\@idrawline[#1](#2,#3){\@ifnextchar ({\@iidrawline[#1](#2,#3)}{\relax}}
\def\@iidrawline[#1](#2,#3)(#4,#5){\@drawline[#1](#2,#3)(#4,#5)
\@idrawline[#1](#4,#5)}
\def\@drawline[#1](#2,#3)(#4,#5){{%
\x@diff=#4\unitlength \advance\x@diff by -#2\unitlength
\y@diff=#5\unitlength \advance\y@diff by -#3\unitlength
\ifx\@linefnt\tenln \linethickness{0.5pt} \else \linethickness{0.9pt}\fi
\lineslope(\x@diff,\y@diff)% returns the two integers in \@xarg & \@yarg.
\@toosmalltrue
{\ifdim\x@diff <\z@ \x@diff=-\x@diff\fi
 \ifdim\y@diff <\z@ \y@diff=-\y@diff\fi
 \ifdim\x@diff >10pt \global\@toosmallfalse\fi
 \ifdim\y@diff >10pt \global\@toosmallfalse\fi}
\@drawitfalse\@horvlinefalse
\ifnum#1 <0 \relax\else\@horvlinetrue\fi
\if@toosmall\@horvlinetrue\fi% to get 'or' condition. We necessarily draw a 
\if@horvline
 \ifdim\x@diff =0pt \put(#2,#3){\ifdim\y@diff >0pt \@linelen=\y@diff \@upline
 				\else\@linelen=-\y@diff \@downline\fi}%
 \else\ifdim\y@diff =0pt
          \ifdim\x@diff >0pt \put(#2,#3){\vrule \@height \@halfwidth \@depth
				\@halfwidth \@width \x@diff}
		\else \put(#4,#5){\vrule \@height \@halfwidth \@depth
				\@halfwidth \@width -\x@diff}\fi
       \else\@drawittrue\fi\fi % construct the line explicitly
\else\@drawittrue\fi
\if@drawit
\ifnum\@xarg< 0 \@negargtrue\else\@negargfalse\fi
\ifnum\@xarg =0 \setbox\@linechar%
\hbox{\hskip -\@halfwidth \vrule \@width \@wholewidth \@height 10.2pt
 \@depth \z@}
\else \ifnum\@yarg =0 \setbox\@linechar%
\hbox{\vrule \@height \@halfwidth \@depth \@halfwidth \@width 10.2pt}
\else \if@negarg \@xarg -\@xarg \@yyarg -\@yarg
        \else \@yyarg \@yarg\fi
\ifnum\@yyarg >0 \@tempcnta\@yyarg \else \@tempcnta -\@yyarg\fi
\setbox\@linechar\hbox{\@linefnt\@getlinechar(\@xarg,\@yyarg)}%
\fi\fi
\if@toosmall% => it isn't a horiz or vert line and is toosmall.
  \@dottedline[\picsquare]{.98\@wholewidth}%
(#2\unitlength,#3\unitlength)(#4\unitlength,#5\unitlength)%
\else
\ifnum\@xarg=0\relax\else\ifdim\x@diff >\z@ \advance\x@diff -\wd\@linechar
  \else\advance\x@diff \wd\@linechar\fi\fi
\ifnum\@yarg=0\relax\else\ifdim\y@diff >\z@\advance\y@diff -\ht\@linechar
  \else\advance\y@diff \ht\@linechar\fi\fi
\ifdim\x@diff <\z@ \@x@diff=-\x@diff \else\@x@diff=\x@diff\fi
\ifdim\y@diff <\z@ \@y@diff=-\y@diff \else\@y@diff=\y@diff\fi
\num@segments=0 \num@segmentsi=0
\ifdim\wd\@linechar >1pt
 \num@segmentsi=\@x@diff \divide\num@segmentsi \wd\@linechar\fi
\ifdim\ht\@linechar >1pt
 \num@segments=\@y@diff \divide\num@segments \ht\@linechar\fi
\ifnum\num@segmentsi >\num@segments \num@segments=\num@segmentsi\fi
\advance\num@segments \@ne %to account for round-off error
\ifnum #1=0 \relax \else\ifnum #1 < -99
  \typeout{***drawline: reduction <= -100 percent implies blankness!***}
\else\num@segmentsi=#1 \advance\num@segmentsi by 100
     \multiply\num@segments \num@segmentsi
     \divide\num@segments by 100
     \ifnum \num@segments=0 \num@segments=1 \fi
\fi\fi
\divide\x@diff \num@segments
\divide\y@diff \num@segments
\advance\num@segments \@ne %for the last segment for which I subtracted
\@xdim=#2\unitlength \@ydim=#3\unitlength
\if@negarg \advance\@xdim -\wd\@linechar\fi
\ifnum\@yarg <0 \advance\@ydim -\ht\@linechar\fi
\@killglue
\loop \ifnum\num@segments > 0
\unskip\raise\@ydim\hbox to\z@{\hskip\@xdim \copy\@linechar\hss}%
\advance\num@segments \m@ne\advance\@xdim\x@diff\advance\@ydim\y@diff%
\repeat
\ignorespaces
\fi%the if of @toosmall
\fi}}% for \if@drawit
\long\def\splittwoargs#1 #2 {(#1,#2)}
\newif\if@stillmore
\newread\@datafile
\long\def\putfile#1#2{\openin\@datafile = #1
\@stillmoretrue
\loop
\ifeof\@datafile\relax\else\read\@datafile to\@dataline\fi
\ifeof\@datafile\@stillmorefalse
\else\ifx\@dataline\@empty \relax
     \else
\expandafter\expandafter\expandafter\put\expandafter\splittwoargs%
\@dataline{#2}
     \fi
\fi
\if@stillmore
\repeat
\closein\@datafile
}

\catcode`\@=12

\usepackage{amsmath,amsfonts,amssymb}%, amstext, amsbsy, amsopn, ,amsthm, }
\renewcommand{\theequation}{\arabic{section}.\arabic{equation}}
\newtheorem{theorem}{Theorem}[section]

\newtheorem{remark}{Remark}[section]

\newtheorem{lemma}{Lemma}[section]

\def\proof{\par \noindent {\it Proof.\/}\quad}

\def\endproof{\quad {\it  QED}\par}
\def\qed{{\it  QED}\par}

\def\sectappend#1
{\vskip35pt plus12pt
{\raggedright
\noindent {\Large \bf {#1}\par}}
\vskip13pt}

\def\build#1_#2^#3{\mathrel{\mathop{\kern 0pt#1}\limits_{#2}^{#3}}}

\newcommand{\biindice}[3]%
{

\begin{array}[t]{c}
#1\\
{\scriptstyle #2}\\
{\scriptstyle #3}
\end{array}

}

\def\beq{\begin{equation}}
\def\eeq{\end{equation}}
\def\beqn{\begin{eqnarray}}
\def\eeqn{\end{eqnarray}}

\def\dis{\displaystyle}

\let\pes\,
\let\esm\:
\let\esn\
\let\ges\;

\def\a{\alpha}
\def\b{\beta}

\def\d{\delta}

\def\ve{\varepsilon}

\def\t{\theta}

\def\r{\rho}

\def\G{\Gamma}
\def\D{\Delta}

\def\L{\Lambda}

\def\naturel{{\mathbb N}}

\def\relatif{{\mathbb Z}}

\let\ZZ=\relatif

\def\reel{{\mathbb R}}
\def\Reel{\mathop{\rm I\! R}}

\newcounter{dessin}
\newenvironment{dessin}{\refstepcounter{dessin} \center}{\endcenter}
\newcommand{\figcaption}[1]{  {\footnotesize Figure \thedessin:  #1}   }
\renewcommand{\thedessin}{\arabic{section}.\arabic{dessin}}

\def\confh{  {\bf h} }
\def\confhh{{\bf \bar h}}

\def\st{\tau}
\def\hm{\bar{h}_{\rm max}}

\def\pfI{Z_1}
\def\pfII{Z_2}
\def\pfIII{Z_3}

\def\feI{\psi_1}
\def\feII{\psi_2}
\def\feIII{\psi_3}

\def\pst{\tau^{\rm pr} }

\def\intu{ \int_{-\frac{u_0}{2}}^{\frac{u_0}{2}} }
\def\limn{  \lim_{N \to \infty}  }
\def\limnb{\lim_{N \to \infty} - \frac{1}{\b N}}

\def\pfe{Z_N}
\def\tf{\Phi^T}
\begin{document}

\title{\Large \bf 
Is there an Optimal Substrate Geometry for Wetting ?
}

												%\spfootnote{Draft \today}  }
\author{\normalsize \sc
                             J. De Coninck,$^{1}$
                             S. Miracle--Sol\'e,$^{2}$
                             and
                           J. Ruiz$^{3}$
               }
\date{{} }
\maketitle
\markboth{\small J. De Coninck, S. Miracle--Sol\'e, 
       and J. Ruiz}{\small Is there an Optimal Substrate Geometry for Wetting ?}
\bibliographystyle{alpha}

\renewcommand{\thefootnote}{}

\footnote{Preprint CPT--99/P.3813, April 1999\\ 
anonymous ftp: ftp.cpt.univ-mrs.fr\\
\noindent www.cpt.univ-mrs.fr}
\renewcommand{\thefootnote}{\arabic{footnote}}
\setcounter{footnote}{0}
\footnotetext[1]{ Centre de Recherche en Mod\'elisation Mol\'eculaire, 
												Universit\'e de Mons--Hainaut, 20 Place du Parc, 			
                                 7000 Mons, Belgium.
                  \hfill\break
                        E-mail address: \textit{joel@gibbs.umh.ac.be\/}  }
\footnotetext[2]{Centre de Physique Th\'eorique, CNRS, Luminy
                                case 907, F-13288 Marseille Cedex 9, France.
                \hfill\break
                         E-mail address: \textit{miracle@cpt.univ-mrs.fr\/} }
\footnotetext[3]{Centre de Physique Th\'eorique, CNRS, Luminy
                              case 907, F-13288 Marseille Cedex 9, France,
                      \hfill\break
                       E-mail address: \textit{ruiz@cpt.univ-mrs.fr\/}  }
\setcounter{footnote}{3}
\thispagestyle{empty}

\footnotesize
\begin{quote}
{\sc Abstract:}
We consider the problem of the Winterbottom's construction and Young's 
equation in the presence of a rough substate and establish their 
microscopic validity within a $1+1$--dimensional SOS type model.
We then  present the  low temperature expansion of the wall tension
leading to the Wenzel's law for the wall tension and its corrections.
Finally, for a fix roughness, we compare the influence of different 
geometries of the substrate on wetting properties. 
We show that 
there is an optimal geometry with a given roughness for a certain 
class of simple substrates.
Our results are in agreement and explain recent numerical simulations.
\\[3pt]
{\sc Key words:} Winterbottom construction, SOS models, Wenzel law, wetting, 
roughness,   interfaces.
%\\[3pt]  {\sc Pacs:} 
\end{quote}
\normalsize
\vskip15pt

\newpage

\baselineskip=14pt
\parskip=12pt plus 1pt minus 1pt
\section{Introduction}
\setcounter{equation}{0}

Wetting phenomena have a long standing history starting with Young 
more than a century ago.
His famous equation describes the behaviour of the contact angle 
$\t$ of a sessile liquid drop $B$ in equilibrium with the vapor phase 
$A$ on top of a substrate $W$:
\begin{equation}
	\label{eq:1.1}
\tau_{AB}\cos\t=\tau_{AW}-\tau_{BW}
\end{equation}
where the $\tau$'s represent the different surface tensions
appearing in the problem.
This equation can be derived for chemically pure substrates
in several ways,
such as by a mechanical argument relative to  the balance of
forces, or by  a thermodynamical argument related to
the minimum of the free energy of the system $ABW$ \cite{A}.
In these approaches, it is in fact 
implicitly assumed that the surface of the substrate 
is perfectly flat.
If this may well be the case at the macroscopic scale
(a few mm), it is far from obvious in the presence of
microroughness.
That is to say, how valid this equation will be on top
of substrates in the presence of atomistic pores or
protusions characteristic of a solid surface ?

To examine this question is precisely the aim of our paper.

Young's equation may be viewed as a direct consequence of the
Winterbottom's construction.
This construction, first obtained from variational principles,
describes the equilibrium shape of 
a crystal as a function of the three different
tensions that appear in the problem.
Its validity 
at the microscopic scale,
together with
the associated contact angle equation,  
has been proved
more recently
in several models \cite{DD, DDR, PV}.
We consider here  this construction 
on top of microscopically rough substrates,
in the case of $1+1$--dimensional
solid-on-solid models.

We use two SOS models in fact:
one to describe the microscopic interface between the
substrate $W$ and the fluids $A$ and $B$,
and another one to describe the microscopic interface between 
$A$ and $B$.
For simplicity, we assume here that the two models have
the same elementary spatial period.

On the other hand, it is known macroscopically that
the roughness of the substrate will induce some 
change in the wall surface tensions, and hence on
the difference $\tau_{AW}-\tau_{BW}$.
This change is described in the literature
by the so-called Wenzel's equation \cite{W}:
$$
\tau_{AW}-\tau_{BW}\quad\hbox{is proportional to}\quad r
$$
where $r$ denotes the ratio between the area $L$ of 
the surface of the substrate and that of its 
projection $L_{0}$ on the tangential plane at the contact point
$$
r=L/L_0
$$
%\newline

We are thus interested to analyze within our model
the validity of this prediction, extending in that way
previous results obtained for the Ising model [BDKZ,BDK].

In particular, we will also be interested by the 
corrections to the Wenzel's law versus the geometry
of the pores or the protusions in our model.
This research clarifies some preliminary results
obtained in that direction with the help of
numerical simulations [TUBD].

The paper is organized as follows.
Section~2 is devoted to the presentation of the model.
Section~3 extends the microscopic validity of the Winterbottom's 
construction to rough substrates.
In Section~4, we present low temperature expansions for the wall 
tension and in Section~5 we compare different geometries for our 
substrate.
Concluding remarks are given in section~6.

\section{The model}
\setcounter{equation}{0}

To define the model, we consider an SOS model where to each site $i$ of the
one dimensional lattice we associate an integer variable $h_i$, $i=0,1,\dots,N$,
which represents the
height of the interface between $i$ and $i+1$.
For a configuration $\confh = \{h_0, \dots,h_N\}$, we draw the 
horizontal lines at height $h_{i}$ between $i$ and $i+1$ 
($i=0,\dots,N-1$), and the vertical lines
at each site $i$, between $h_{i-1}$ and $h_i$. 
We use $\G$ to denote  the corresponding polygonal line 
(see Fig.~\ref{F2.1}).
Its length is
$|\G| = \sum _{i=1}^N ( 1 + |h_i - h_{i-1}|)$.

We want here to study this interface  on top of  a \textit{rough substrate\/} with roughness $r$.
 The substrate is thus represented in our case 
 by a periodic SOS interface $W$, with periodicity $a$, and height configuration
$\confhh =\bar{h}_0,\dots,\bar{h}_N$ where 
$\bar{h}_i = \bar{h}_{a+i}$, so that  
$$
r=1+ \dis \frac{ \sum_{i=1}^{a} | \bar{h}_i - \bar{h}_{i-1}| }{a}
$$

%\begin{quote} \it
%On pourrait consid\'erer une forme plus g\'en\'erale avec
%la condition moins restrictive
%$$\bar{h}_{a+i+1} - \bar{h}_{a+i}= \bar{h}_{i+1} - \bar{h}_{i}$$
%Ceci introduit un substrat avec un angle $\bar{\theta}$ : 
%$\bar{y} = \tan \bar{\theta} = \bar{h}_a -\bar{h}_0$
%\rm
%\end{quote}

The energy of a configuration, in a box of length $N$
(which will be taken as a multiple of $a$), is given by
\begin{equation}
H_N(\G, W)
=
J_{AB} | \G  \setminus (\G \cap W) |
+ J_{AW} |  \G \cap W |
+J_{BW} | W  \setminus (\G \cap W) |
\label{eq:2.1}
\end{equation}
Here $\G$ is above $W$, which means $h_i \geq \bar{h}_i$ for all $i$.
The set  $\G\setminus(\G\cap W)$ is relative to the $AB$
microscopic interface, $\G\cap W$ defines the part of the
substrate in contact with   $A$, and
$W\setminus(\G\cap W)$ is relative to the contact zone between
$B$ and $W$.

This system describes a system of droplets of a phase $B$ inside 
a medium $A$ on top of the wall $W$.
$J_{AB}$, $J_{AW}$ and $J_{BW}$ are the energies per unit length of 
the corresponding microscopic interfaces
(see Fig.~\ref{F2.1}).

\begin{center}
\setlength{\unitlength}{10mm}
\begin{picture}(11,7)
    \put(0,-2){
\begin{picture}(11,10)
    %exitation d0
\thinlines
 \drawline(0,2)(11,2)
 \drawline(11,2)(11,9)
\drawline(11,9)(0,9)
\drawline(0,2)(0,9)

%le mur
\thicklines

\drawline(0,4)(2,4)
\drawline(2,4)(2,3.5)
\drawline(2,3.5)(3,3.5)
\drawline(3,3.5)(3,3)
\drawline(3,3)(5,3)
\drawline(5,3)(5,4)

\drawline(5,4)(7,4)
\drawline(7,4)(7,3.5)
\drawline(7,3.5)(8,3.5)
\drawline(8,3.5)(8,3)
\drawline(8,3)(10,3)
\drawline(10,3)(10,4)

\drawline(10,4)(11,4)

%le contour

\drawline(0,4.1)(2.1,4.1)
\drawline(2.1,4.1)(2.1,3.6)
\drawline(2.1,3.6)(3.1,3.6)
\drawline(3.1,3.6)(3.1,3.1)
\drawline(3.1,3.1)(3.6,3.1)
\drawline(3.6,3.1)(3.6,7.1)
\drawline(3.6,7.1)(4.1,7.1)
\drawline(4.1,7.1)(4.1,7.6)
\drawline(4.1,7.6)(5.6,7.6)
\drawline(5.6,7.6)(5.6,7.1)
\drawline(5.6,7.1)(6.1,7.1)
\drawline(6.1,7.1)(6.1,4.1)
\drawline(6.1,4.1)(6.6,4.1)
\drawline(6.6,4.1)(6.6,5.1)
\drawline(6.6,5.1)(7.6,5.1)
\drawline(7.6,5.1)(7.6,6.6)
\drawline(7.6,6.6)(9.6,6.6)
\drawline(9.6,6.6)(9.6,3.6)
\drawline(9.6,3.6)(9.9,3.6)
\drawline(9.9,3.6)(9.9,4.1)
\drawline(9.9,4.1)(11,4.1)

\put(1.5,8){$A$}
\put(4.5,5.5){$B$}
\put(6.6,3.1){$W$}
\put(6.3,7.1){$\Gamma$}

\put(.7,4.3){$J_{AW}$}
\put(4.7,7.8){$J_{AB}$}
\put(4,3.2){$J_{BW}$}
\end{picture}
}
\end{picture}
\end{center}

\begin{dessin}
\label{F2.1} 
\figcaption{A configuration of the interface $\G$ on the substrate $W$.}
\end{dessin}

Let us first introduce the different tensions appearing in the problem.

The surface tensions associated to the macroscopic interfaces $AB$ 
and $AW$ are defined as follows :
\begin{equation}
\st_{AB}(\t) =
\limn
-\dis \frac{\cos \t}{\b N}
\ \log \sideset{}{^{*}} \sum_{\G} 
\exp (-\b J_{AB} | \G | )
\label{eq:2.2}
\end{equation}
where the sum $\sum^{*}$ runs over all configurations satisfying 
$h_0 = 0$ and $h_N = N \tan \t$, and
\begin{equation}
\dis
\st_{AW} =
\limnb
\ \log \sideset{}{^{\dag}} \sum_{\G} 
\exp [-\b H_N(\G  ,W) ]
\label{eq:2.3}
\end{equation}
where the sum  $\sum^{\dag}$ runs over all configurations such that 
$h_0 = \bar{h}_0$ and $h_N = \bar{h}_N$.
Finally, for the interface $BW$, we have 
\begin{equation}
\st_{BW} = r J_{BW}
\label{eq:2.4}
\end{equation}

Let us point out that the anisotropy of the SOS model considered here 
leads to an orientation dependent surface tension for the $AB$ 
interface. That the limits exist follows from standard arguments, see 
e.g.\ \cite{DD,MMR}.

\section{Winterbottom's  construction  on  rough \\ substrate}
\setcounter{equation}{0}

To analyze the microscopic problem of the Winterbottom's construction for 
the model under consideration, we have to consider, following 
\cite{DDR}, the system given by the Hamiltonian $H_{N}$
submitted to a canonical constraint on the volume of phase $B$
enclosed between the $AB$ and $W$ interfaces.
The first step consists to analyze the case of a single droplet of 
the phase $B$.
To this end, following \cite{MR}, we consider the  Gibbs ensemble consisting
of  the configurations which have  specified height  at extremities
and which have a specified volume $V$
between the interface $AB$ and the substrate $W$,  
\begin{equation}
h_N = \bar{h}_N, \quad
h_0 =\bar{h}_0+ M, \quad
\sum _{i=0}^N  ( h_i - \bar{h}_i ) =V,
\label{eq:3.1} 
\end{equation}
We assume that the constraints due to the fact that the 
microscopic interface does not touch the substrate
\begin{equation}
h_i>\bar{h}_i , \quad i=0,\dots, N
\label{eq:3.2}
\end{equation}
are satisfied, therefore there is no interaction between the 
interface and the substrate. 
The corresponding partition function is
\begin{equation}
\begin{array}{ll}
\dis 
\pfI (N,V,M,\confhh) = \sum_{\confh} e^{- \beta H_{N}({\confh})}
\ \delta (h_N-\bar{h}_N) \delta (h_0 - \bar{h}_0-M)
&
\\
\hphantom{xxxxxxxxxxxxxxxxxxxxxx}
\dis
\delta (\sum _{i=0}^N  ( h_i - \bar{h}_i ) - V)
\prod_{i=1}^{N} \chi (h_i > \bar{h}_i)
&
\end{array}
\label{eq:3.3} 
\end{equation}
where
\begin{equation} 
H_N ({\confh}) = \sum _{i=1}^N J_{AB}( 1 + |h_i - h_{i-1}|) 
\label{eq:3.4} 
\end{equation}
Hereafter, $V$ and $M$ must be understood
as their integer part when they do not belong to $\relatif$.

\noindent
We denote by 
$\pfI (N, V , M )$,
the sum of the same Boltzmann factors
$e^{- \beta H_N({\confh})}$
over the configurations satisfying conditions 
(\ref{eq:3.1}) and 
\begin{equation}
h_i \geq 0 , \quad i= 0, \dots , N
\label{eq:3.5}
\end{equation}

We next consider a conjugate ensemble with the partition function:
\begin{equation} 
\pfII (N,u,\mu) = \sum _{\confh} e^{- \beta H_N({\confh})}
\ e^{\beta u (V(\confh) / N) + \beta \mu  h_0 }
\ \delta (h_N) 
\label{eq:3.6} 
\end{equation}
where $V(\confh)=\sum_{i=0}^{N} h_i$ and  $u \in \reel $ and $\mu \in \reel $
are the conjugate variables to $M$ and $V$ in $Z_{1}$.

Our first Theorem establishes the existence of the
thermodynamic limit for these ensembles and their equivalence
in this limit.

\begin{theorem}
 The following limits exist
\beqn 
\feI (v,m) 
&=& 
\limnb
\  \ln \pfI (N, v N^2 , mN, \confhh) 
\label{eq:3.7} 
\\
&=& 
\limnb
\  \ln \pfI (N, v N^2 , mN ) 
\label{eq:3.8}
\\
\feII (u,\mu) 
&=& 
\limnb
\  \ln \pfII (N, u,\mu ) 
\label{eq:3.9} 
\eeqn
They define the free energies per site associated to the considered 
ensembles as, respectively,  
convex and concave functions of their variables.
Moreover, $\feI$ and $- \feII$ are conjugate convex functions:
\begin{equation}
\begin{array}{rcl}
- \feII (u,\mu)  
&=&
\displaystyle \sup_{v,m} \  [ u v + \mu m - \feI (v,m) ] 
\\
 \feI (v, m) 
&=&
\displaystyle \sup_{u,\mu} \  [ u v + \mu m + \feII (u,\mu) ] 
\end{array}
\label{eq:3.10}
\end{equation}
\end{theorem}

\proof
We take, for simplicity of the notations,
$\bar{h}_N =0$
and let 
$\hm =\max_i |\bar{h}_i |$.
Then 
\begin{equation}
\pfI (N,V_-,M) e^{-2 \b |\hm |}
\leq
\pfI (N,V,M, \confhh)
\leq
\pfI (N,V_+,M) e^{2 \b |\hm|}
\label{eq:3.11}
\end{equation}
where 
$$
V_{\pm} = V \pm (N-1) \hm
$$
These inequalities follow by using respectively 
the changes of variables 
$$
\begin{array}{lll}
h_0 &\to & \tilde{h}_0 = h_0
\\
h_N &\to & \tilde{h}_N = h_N
\\
h_i &\to & \tilde{h}_i = h_i \pm \hm , \quad i=1, \dots, N-1
\end{array}
$$
Now we use the subaddittivity property
\begin{equation} 
\pfI ( 2N, 2(V' + V''), M' + M'' ) \ge 
\pfI ( N, V', M')\  \pfI ( N, V'', M'') \  
e^{- 2 \beta \frac{|M''|}{2N-1} }  
\label{eq:3.12}
\end{equation}
The proof of this property is given in \cite{MR} 
and we recall it in the appendix for the reader's convenience.
From this property, we get the existence of the limits 
when $N\to\infty$ of
$$
(-1/\b N) \log \pfI (N,V_- ,mN)
\hbox{ and }
(-1/\b N) \log \pfI (N,V_+ ,mN)
$$
and the convexity of the corresponding free energy
(we take $N=2^n$, $n \in \naturel$).
Provided that $\hm =o(N)$ (in fact $\hm$ is a constant 
under the hypothesis of section 2)
we see from inequality (\ref{eq:3.11}) that 
these limits actually coincide with the limit
$$
\feI (v,m) 
= 
\limnb
\  \ln \pfI (N, v N^2 , mN,\confhh ) 
$$
which is thus independent of $\confhh$, and where
$$
v= \limn \frac{V_+}{N^2}=\lim_{N\to \infty} \frac{V_-}{N^2},
\quad
m= \limn \frac{M}{N}
$$
We next introduce the partition functions
\beqn
\pfII^+(N,u,\mu)
&=&
\sup_{V,M \in \relatif } 
\left[ e^{\beta u  (V/N) + \beta \mu M } 
\  \pfI(N,V,M)
 \right]
 \nonumber 
 \\ 
 \widetilde{\pfII}(N,u,\mu)
&=&
\sum_{V,M \in \relatif } 
\left[ e^{\beta u  (V/N) + \beta \mu M } 
\  \pfI(N,V,M)
 \right]
 \nonumber
\eeqn
and the convex function
$
\feII^{\ast} (u,\mu) = \sup_{v,m} \  [uv+\mu m -\feI (v,m)] 
$. 
The Griffiths maximum principle adapted to our case (see \cite{MR}\cite{GMS})
gives that  
$$\limnb \log \pfII^+(N,u,\mu)
=
\limnb \log \widetilde{\pfII}(N,u,\mu)
=  
- \feII^{\ast} (u,\mu)
$$
These limits coincide with the limit of
$ (-1/\b N) \log {\pfII}(N,u,\mu)$, 
because
$\widetilde{\pfII}(N,u,\mu) / \pfII (N,u,\mu)\to \xi > 0$
as shown in  \cite{DDR}.
\qed

We have thus shown that the free energy of the sessile drop with the 
fixed volume $V= vN^2$ is proportional to $N$.
Using \eqref{eq:3.10}, we can determine this free energy using the 
conjugate ensemble where the constraints on the volume and on the 
solid surface do not appear.
We shall be interested in the case $m=0$ 
which corresponds to a drop on the horizontal plane.
We write
$$
\feI(v) = \feI(v,m=0),\quad \feII(u)= \feII(u, \mu =- u/2) 
$$
and will express these free energies in terms of the surface tension
$\st_{AB}$.
We introduce for that the projected surface tension
$$
\pst (-\tan \theta) = \st_{AB}(\t) / \cos \t
$$
and its Legendre transform %$\vp$ 
\begin{equation}
- \varphi (x) 
=
\displaystyle \sup_y \  [ x y - \pst (y) ] 
\label{eq:3.13}
\end{equation}
According to Andreev \cite{An},  the Legendre transform $\varphi$ 
solves the Wulff variational problem when the 
surface tension  is $\st_{AB}(\t)$.
The graph of $\varphi$ gives the boundary of the 
equilibrium crystal shape of the phase $B$ inside the phase $A$. 
In the case under consideration, this Legendre transform 
corresponds to the free energy associated with the statistical ensemble 
conjugate to the  ensemble defining $\st_{AB}(\t)$
with respect to the constraint on $h_{N}$ 
\cite{MR} (see \eqref{eq:3.23} below).
The value of $\varphi$ is given by 
$$\varphi (x) = -\frac{1}{\b} 
\log \frac{e^{\b J_{AB}} \cosh \b J_{AB}}{\cosh \b J_{AB}-\cosh \b x }$$

The following theorem establishes the microscopic validity of the 
associated
Wulff's construction.

\begin{theorem}
	\label{thm3.2}
The free energies $\feI$ and $\feII$
can be expressed in terms of the functions
$\varphi$ and $\pst$
as follows
\beqn 
\feII (u) 
&=&
\frac{1}{u} \int _{-\frac{u}{2}} ^{\frac{u}{2}}  \varphi (x ) dx 
\label{eq:3.14} 
\\
\feI (v) 
&=& 
\frac{2}{u_0} 
\intu \varphi (x) dx 
-  \varphi (  u_0 /2)
\label{eq:3.15} 
\\
&=&
\frac{1}{u_0}    
\intu  \pst (\varphi' (x)) dx 
\label{eq:3.16}
\eeqn
where 
$u_0$  satisfies
\begin{equation}
 \frac{1 }{ {u_0^2} } 
\intu \varphi (x) dx 
- \frac{1 }{ {u_0} } \varphi (  u_0 /2) = v 
\label{eq:3.17}
\end{equation}
\end{theorem}

\proof%of{Theorem \ref{thm3.2}}
Consider the Legendre relation (\ref{eq:3.10}).
The supremum over $u,\mu$ is obtained for the value $ u_0, \mu_0 $
for which the partial derivatives of the right hand side are zero:
$$ 
v + (\partial \feII/ \partial u)  ( u_0,\mu_0 ) = 0, \; 
m + (\partial \feII/ \partial \mu)  ( u_0,\mu_0 ) = 0  
$$ 
That is, for $ u_0, \mu_0 $  which satisfy 
\beqn
 \frac{1 }{ {u_0^2} } 
\int _0 ^{u_0} \varphi (x+ \mu_0) dx 
- \frac{1 }{ {u_0} } \varphi ( \mu_0 + u_0 ) 
&=&
v
\label{eq:3.18}
\\
 \frac{1 }{ u_0 } 
[\varphi (\mu_0 ) - \varphi ( \mu_0 + u_0) ]
&=&
m
\label{eq:3.19}
\eeqn
The function $\varphi$ is even and thus  equation (\ref{eq:3.19}) for $m=0$
gives   $\mu_0 =u_0 / 2$; 
inserting this value in (\ref{eq:3.18}) gives
expression (\ref{eq:3.17}) in the Theorem, and the Legendre relation (\ref{eq:3.10})
reads now
\begin{equation}
\feI (v) = u_0 v + \feII (u_0)
\label{eq:3.20}
\end{equation}
if 
$ \feI'  (v ) = u_0$
or 
$ \feII'  ( u_0) = -v   $. 
The proof now proceeds as the proof of Theorem~3 in  \cite{MR}
whose main ingredient is the following computation of $\feI (v)$.
We consider  the difference variables
\begin{equation} 
n_i = h_{i-1} - h_i, \quad i = 1,...,N  
\end{equation}
and observe that
$
 V(\confh)= \sum _{i=0}^N h_i 
= \sum _{i=1}^N \; i n _i 
$.
Thus
\begin{equation} 
\pfII (N,u,\mu) 
= \prod _{i=1} ^N \Big( \sum _{n_i \in \relatif }
e^{ - \beta J_{AB}(1+ |n_i|) + \beta ( u / N ) i n_i + \beta \mu n_i} \Big) 
\end{equation}
Since, on the other hand (see \cite{MR})
\begin{equation}
\varphi (x) = 
\limnb
\log
 \sum _{\confh} e^{- \beta H_N({\confh})+\b x h_N} \d(h_0)
= 
- \frac{1}{\beta} \ln \sum _{n\in \relatif } 
e^{ - \beta J_{AB}(1+|n|) + \beta x n} 
\label{eq:3.23} 
\end{equation}
we have
$$
\pfII (N,u, \mu ) 
= \exp \Big( - \beta \sum _{i=1} ^N
\varphi \big( \frac{ u }{ N} i + \mu \big) \Big) 
$$
and
\beqn
\feII (u,\mu)
& = &
\limn \frac{1 }{ N} 
\sum _{i=1}^N \varphi \big( \frac{ u }{ N} i + \mu \big) 
=
\limn \frac{1 }{ u} 
\sum _{i=1}^N \frac{u }{ N} 
\varphi \big( \frac{ u }{ N} i + \mu \big)
\label{eq:3.24}
\\ 
&=&
\frac{1 }{ u} \int _0 ^u \varphi (x + \mu) dx 
\eeqn
which for $\mu = - u/2$ gives expression (\ref{eq:3.14}). The theorem
then follows by using the Legendre transform relations.
\endproof

From Theorem~\ref{thm3.2} we know that the equilibrium shape is a 
piece of the associated Wulff shape. To determine which piece, let us 
introduce the ensemble with   partition function
\begin{equation}
\pfIII(V,\D \tau ) 
=
\sum_N e^{\b \D \tau N} \pfI(N,V,M=0)
\label{eq:3.26}
\end{equation}
This ensemble is the conjugate ensemble of the ensemble associated to 
partition function $Z_{1}$ with respect to the variable $N$, the 
conjugate variable being the difference of wall tensions
$\D \tau = \tau_{AW} - \tau_{BW}$.
\begin{theorem}
The limit 
\begin{equation}
\feIII (\D \tau) =
\lim_{V \to \infty} - \frac{1}{\b \sqrt{V}} \log \pfIII(V,\D \tau ) 
\label{eq:3.27}
\end{equation}
exits and 
\begin{equation}
\feIII (\D \tau) 
=
- \sup_{\a}
[  \a \D \st - \a \feI (\a^{-2}) ]
\label{eq:3.28}
\end{equation}
Moreover
\begin{equation}
\feIII (\D \tau) 
=
2\left[ 
\intu \varphi (x) dx  - u_0 \varphi ( u_0 /2 )
\right]^{\frac{1}{2}}
\label{eq:3.29}
\end{equation}
where $u_0$ is the solution of  
\begin{equation} 
\D \tau = \varphi (u_0/2)
\label{eq:3.30}
\end{equation}
\end{theorem}

\proof
The partition function $\pfIII$ corresponds to the conjugate
Gibbs ensemble of the ensemble associated to $\pfI$.
Equation (\ref{eq:3.28}) is the expression of this fact for the 
corresponding free energies.
This can be seen from
$$
\pfIII (\D \tau) 
\simeq 
\sum_N e^{\b \D \tau N     
- N \feI (\frac{V}{N^2})}
\simeq 
\sup_N e^{-\b\sqrt{V}[-\D \tau \frac{N}{\sqrt{V}} 
+ \frac{N}{\sqrt{V}} \feI (\frac{V}{N^2})]}
$$
which, by taking $\a={N}/{\sqrt{V}}$, implies equation (\ref{eq:3.28}).
A rigorous proof of equation (\ref{eq:3.28})  may be obtained,
as in Theorem 3.1, 
by the Griffiths maximum principle (see \cite{GMS}).

In order to compute $\feIII (\D \tau)$ we first remark that
from (\ref{eq:3.28}) we get
\begin{equation}
\feIII (\D \tau) = -\a_0 \D \st + \a_0 \feI (\a_0^{-2})
\label{eq:3.31}
\end{equation}
where $\a_0$ is the solution of
\begin{equation}
\frac{\partial}{\partial\a}(-\a \D \st + \a \feI (\a^{-2}))
= - \D \st + \feI (\a^{-2}) -2\a^{-2} \feI'(\a^{-2}) = 0
\label{eq:3.32}
\end{equation}
The function $\feI$ is given by formula (\ref{eq:3.15}) which according
to equation (\ref{eq:3.17}) can be written as
\begin{equation}
\feI (v)=2u_0 v + \varphi(u_0/2)
\label{eq:3.33}
\end{equation}
where $u_0$ is the solution of \eqref{eq:3.17}.
Deriving with respect to $v$ equation \eqref{eq:3.20} we find
\begin{equation}
\feI' (v)=u_0
\label{eq:3.34}
\end{equation}
Taking $\a^{-2}=v$, equation (\ref{eq:3.32}) reads
$
\D \tau + \feI(v) - 2v\feI'(v) = 0
$
which, taking (\ref{eq:3.33}) and (\ref{eq:3.34}) into account, amounts to 
\eqref{eq:3.30} :
% \begin{equation}
$ \D \tau - \varphi(u_0/2) = 0$.
Let $v_0$ be the value of $v$ which according to \eqref{eq:3.17} corresponds
to the value of $u_0$ which solves the previous equation, that is
\begin{equation}
v_0 = \frac{1 }{ {u_0^2} } 
\intu \varphi (x) dx 
- \frac{1 }{ {u_0} } \varphi (\frac{{u_0}}{ 2})
\label{eq:3.35}
\end{equation}
Then from (\ref{eq:3.31}) we get
$
\feIII (\D \tau) = -\frac{\D\tau}{\sqrt{v_0}} + 
\frac{1}{\sqrt{v_0}} \feI (v_0)
$
and, using (\ref{eq:3.33}) and (\ref{eq:3.30}),
$$
\feIII (\D \tau) = -\frac{\D\tau}{\sqrt{v_0}} + 
\frac{1}{\sqrt{v_0}} (2u_0v_0 + \varphi (\frac{{u_0}}{ 2}))
= 2u_0\sqrt{v_0}
$$
This proves equation (\ref{eq:3.29}) stated in the
theorem.
\endproof

Theorem~3.3 gives, by \eqref{eq:3.30}, the contact angle relation 
(see e.g.\ Remarks~6 and 8 in \cite {MMR})
\begin{equation}
	\st_{AB}(\t) \cos \t - \st_{AB}' (\t) \sin \t
	= \st_{AW}-\st_{BW}
	\label{eq:3.36}
\end{equation}This relation reduces itself to Young's equation 
\eqref{eq:1.1} for 
isotropic media.

\begin{remark}\label{rem.3.1}
	\rm
More general Hamiltonians of the form
$H=\sum P(|h_i - h_{i-1}|)$,
where $P$ is a strictly increasing function such that
$P(x)\geq |x|$, when $x\to \infty$, can be treated in the same way
as well as
%The proofs extend also to 
the case of continuous height variables.
\end{remark}

\begin{remark}\label{rem.3.2}
	\rm
	Let us also stress that the proofs may be easily extended to the cases 
of finite range interactions between the interface $AB$ and the 
substrate $W$.
\end{remark}

Theorem~3.3 establishes the validity of the Winterbottom's construction 
for  a single droplet.
Actually the initial problem of an interface attracted by a substrate with a 
given volume of the phase $B$  leads to the analysis of a gas of 
droplets
with a global volume constraint
(see \cite{DDR} section~4).
This problem has been considered in \cite{DDR} and 
it has been proven there (in the case of a flat substrate) that 
for a large set of configurations, whose probability tends to one 
when $N\to \infty$, there 
is only one large droplet, which has the macroscopic volume 
given by the constraint  and a 
gas of microscopic droplets without any volume constraint.
The free energy of this gas  is given by the wall 
surface tension.
This shows that the free energy of the large droplet is given by 
Theorem~3.3 and justifies the way we tackle the problem provided the same 
analysis is extended to the case of rough substrate. We will not give 
here the details of such analysis which can be easily adapted from 
section~4 in \cite{DDR}.

 \section{Low temperature expansion of the wall 
%\hfill\break medium surface 
tension}
\setcounter{equation}{0}
This section is devoted to study the behavior, at low temperatures
of the surface tension $\st_{AW}$, defined by equation (\ref{eq:2.3}).

Consider a drop of $B$ on the top of the substrate $W$.
Two cases may appear: either the liquid $B$ is always in contact with 
$W$ 
or there may be droplets of $A$ between the liquid and $W$.
Within our SOS model, that means that the ground state of the 
Hamiltonian of the system is given by the microscopic interface $\G$
that coincides with the substrate $W$, or microscopic interface $\G$ 
which leave 
holes between $\G$ and  $W$.

Let us  develop the first case.
To this end, we introduce the energy difference
\begin{equation}
H'_N(\G , W) =H_N(\G,W) - H_N(W, W)
\end{equation}
so that the surface tension $\st_{AW}$ reads
\begin{equation}
\st_{AW} = r J_{AW} + \limnb \log \pfe
\label{eq:4.2}
\end{equation}
where
\begin{equation}
\pfe = \sum_{\G} e^{- \b H'_N (\G, W )}
\end{equation}
Our first step is to write $\pfe$ as the partition function of a gas of
elementary excitations, simply also called excitations, which can
be viewed as microscopic droplets over the substrate. 
These excitations are defined as follows.
Given $\G$ and $W$, we consider the symmetric difference
\begin{equation}
\D = (\G \cup W) \setminus (\G \cap W)
\label{eq:4.4}
\end{equation}
We decompose $\D$ in maximal connected components
$\D = \d_1\cup \d_2\cup \dots\cup\d_n$ called excitations.
Here, two components are said connected if they are connected considered as 
subsets of $\reel^2$.
A set $\{ \d_1,\d_2,\dots,\d_n \}$ of mutually disjoint excitations is
called an {\it admissible\/} family of excitations.
Then there exists a microscopic interface (SOS configuration) $\G$,
such that $\D = \d_1\cup \d_2\cup \dots\cup\d_n =(\G \cup W)\setminus 
(\G \cap W)$.
It is obtained by the formula
\begin{equation}
\G = (\D \cup W) \setminus (\D \cap W)
\label{eq:4.5}
\end{equation}
This correspondence between admissible family of excitations
and interfaces SOS configurations is one-to-one.

The energy difference 
$H'_N$ reads in terms of families of excitations as 
$$
H'_N(\G,W)
=
E(\d_1)+\dots+E(\d_n)
$$
where 
\begin{equation}
E (\d) = J_{AB} | \d \setminus (\d \cap W) | 
- (J_{AW}-J_{BW}) |  (\d \cap W) |
\label{eq:4.6}
\end{equation}
Indeed
\beqn 
H'_N
&=&
J_{AB} | \G \setminus (\G \cap W) | 
+ J_{AW} | \G \cap W | 
+ J_{BW} | W \setminus (\G \cap W) | 
- J_{AW} |W|
\nonumber
\\
&=&
J_{AB} | \G \setminus (\G \cap W)| 
- (J_{AW}-J_{BW}) | W \setminus (\G \cap W) | 
\nonumber
\\
&=&
J_{AB} | \D \setminus (\D \cap W) | 
- (J_{AW}-J_{BW}) |  \D \cap W |
\nonumber 
\eeqn
The equality 
$\G \setminus (\G \cap W) = \D \setminus (\D \cap W)$ 
follows from
equations (\ref{eq:4.4}), (\ref{eq:4.5}) and 
$\G \cup W = \D \cup W$.
The equality 
$| W \setminus (\G \cap W) |= |\D \cap W|$ 
follows from
equation  (\ref{eq:4.5}) and the relations
$\G \cup W = \D \cup W$,  
$|W| +|\G| = |W \cup \G| + |W \cap  \G | $.

%We introduce the infinite cylinder
%$\L_j = \{ x=(x_1, , x_2) \in \reel^2 : 0 \leq x_1 \leq j \}$, 
Then 
\begin{equation}
\pfe
=
\sum_{\D = \{\d_1,\dots, \d_n\}\subset \L_N}
\prod_{i=1}^{n} e^{-\b E(\d_i)}
\end{equation}
where the sum runs over admissible families of excitations whose 
projection is included in the infinite cylinder
$\L_N = \{ x=(x_1,  x_2) \in \Reel^2 : 0 \leq x_1 \leq N \}$ 
and the product
is taken equal to 1 if $\D=\emptyset$.

In the concept of excitation that we are
considering, the configuration $\G=W$,
in which the microscopic interface is following the wall, 
as the ground state of the system.
In other words, we assume that
$H'_N(\G,W)>0$ for all $\G$ and $N$, or equivalently, that
\begin{equation}
\min_\d E(\d)> 0
\label{eq:4.8}
\end{equation}
In fact it is enough that this condition is satisfied for $N=a$,
that is for all excitations belonging to $\L_a$.

We next consider arbitrary families of elementary excitations 
non necessarily mutually compatible and in which a given excitation 
can appear several times.
To any such family $\{\d_1,\dots, \d_n\}$ a graph $G(d_1,\dots, \d_n)$
is associated in such a way that to each excitation corresponds 
(in a one-to-one way) a vertex of the graph, and
there is an edge joining the vertex corresponding to $\d_i$ and $\d_j$
whenever $\d_i$ and $\d_j$ are not compatible or coincide.
We introduce the {\it clusters\/} $C$ as the  arbitrary families of 
excitations for which the associated graph $G(d_1,\dots, \d_n)$
is connected (this means that the excitations draw a connected set
in $\Reel^2$).
Then we get
\begin{equation}
\log \pfe = \sum_{C} \tf (C)
\label{eq:4.9}
\end{equation}
where the sum runs over all clusters whose excitations belong to $\L_N$.
The truncated functions $\tf$ are defined by
\begin{equation}
\tf (\d_1,\dots,\d_n) = 
\frac{a(\d_1,\dots,\d_n)}{n!}\prod_{i=1}^{n} e^{-\b E(\d_i)}
\label{eq:4.10}
\end{equation}
%$$
%E(\d) > {\rm const}\  |\d|
%$$
the arithmetic coefficient being

\begin{equation}
	a(\d_1,\dots,\d_n)=\sum_{G\subset G(\d_1,\dots,\d_n)}(-1)^{\ell(G)}
	\label{eq:4.11}
\end{equation}
Here the sum runs over all connected subgraphs $G$ of $G(\d_1,\dots,\d_n)$,
whose vertex coincide with the vertex of $G(\d_1,\dots,\d_n)$,
and $\ell(G)$ is the number of edges of the graph $G$.
If the cluster $C$ contains only one excitation then $a(\d)=1$. 

To express condition (\ref{eq:4.8}) in terms of the coupling constants,
we need a description of the substrate.
Let $\G(z)$ be the horizontal line at height $z$, that is 
$h_i =z$ for all  $i$.
For any $z\in \ZZ$ such that $\inf_i \bar{h}_i +1\leq z \leq \sup_i \bar{h}_i$,
the substrate $W$ and the line $\G(z-\ve)$, $0< \ve < 1$, 
intersect in a finite 
number of points, 
$W \cap \G(z-\ve) = \{ A_1,A_2, \dots,A_p\}$,
ordered in such a way that the first coordinates
$i_k$ 
($k=1,\dots,p$) of $A_k$
satisfy
$i_1 < i_2 < \cdots <i_p$.
The part of $W$ between the two points
$B_k=(i_k,z)$ and $B_{k+1}=(i_{k+1},z)$ 
lies either below either above or on the
the substrate $W$.
It is called a well in the first case and we denote it by
$w_k(z)$,  and a protusion in the second case (see Fig.\ \ref{F4.1}).
We let 
$\r = \max_{z,k} |w_k(z)| /( i_{k+1}-i_k )
=\max_{z,k} | \d_k(z)\cap W | / |\d_k(z) \setminus  (\d_k(z)\cap W) |$,
where $ \d_k(z)$ is the excitation 
$\d_k(z)= w_k(z) \cup [i_k,i_{k+1}]$.
\begin{center}
\setlength{\unitlength}{9mm}
\begin{picture}(11,10)
\linethickness{0.15mm}
\put(0,0){\line(0,1){10}}
\put(11,0){\line(0,1){10}}
\put(0,0){\line(1,0){11}}
\put(0,10){\line(1,0){11}}
\linethickness{0.25mm}
\put(0,5){\line(1,0){1}}
\put(1,5){\line(0,1){1}}
\put(1,6){\line(1,0){1}}
\put(2,6){\line(0,-1){1}}
\linethickness{0.5mm}
\put(2,5){\line(0,-1){2}}
\put(2,3){\line(1,0){1}}
\put(3,3){\line(0,-1){2}}
\put(3,1){\line(1,0){1}}
\put(4,1){\line(0,1){4}}
\linethickness{0.25mm}
\put(4,5){\line(0,1){4}}
\put(4,9){\line(1,0){2}}
\put(6,9){\line(0,-1){2}}
\put(6,7){\line(1,0){1}}
\put(7,7){\line(0,-1){1}}
\put(7,6){\line(1,0){1}}
\put(8,6){\line(0,-1){1}}
\linethickness{0.5mm}
\put(8,5){\line(0,-1){3}}
\put(8,2){\line(1,0){1}}
\put(9,2){\line(0,1){2}}
\put(9,4){\line(1,0){1}}
\put(10,4){\line(0,1){1}}
\linethickness{0.25mm}
\put(10,5){\line(1,0){1}}
\put(0,4.8){\line(1,0){11}}
\multiput(2,2.9)(0,-0.2){15}{\line(0,-1){0.1}}
\multiput(4,0.9)(0,-0.2){5}{\line(0,-1){0.1}}
\multiput(8,1.9)(0,-0.2){10}{\line(0,-1){0.1}}
\multiput(10,3.9)(0,-0.2){20}{\line(0,-1){0.1}}
\put(0.6,6.2){$W$}
\put(1.4,5){$B_1$}
\put(3.4,5){$B_2$}
\put(7.4,5){$B_3$}
\put(9.4,5){$B_4$}
\put(1.45,0.2){$i_1$}
\put(3.45,0.2){$i_2$}
\put(7.45,0.2){$i_3$}
\put(9.45,0.2){$i_4$}
\put(1.84,4.9){$\times$}
\put(3.84,4.9){$\times$}
\put(7.84,4.9){$\times$}
\put(9.84,4.9){$\times$}
\put(0.1,4.3){$\Gamma ( z- \varepsilon)$}
\put(3.2,3){$w_1$}
\put(8.3,4){$w_3$}
\end{picture}
\end{center}
\begin{dessin}
\label{F4.1} 
\figcaption{The wells $w_1$ between $B_1$ and $B_2$ and $w_3$ between
$B_3$ and $B_4$. }
\end{dessin}
Then condition (\ref{eq:4.8}) reads
\begin{equation}
J_{AB}>\r (J_{AW} - J_{BW})
\label{eq:4.12}
\end{equation}
\def\wall{W}

Hereafter it will be more convenient to denote $\wall$ the infinite 
periodic wall whose restriction to $\L_{N}$ is given by the previous 
height $\bar{h}_{0},\dots,\bar{h}_{N}$.
Notice that the expression \eqref{eq:4.6} of the energy of excitation 
remains unchanged.
%We introduce  $\wall$ as the infinite periodic wall 
%whose restriction to $\L_N$ is $W$ 
We shall use $W_{a}$ to denote the restriction of $W$ to $\L_a$.

	\begin{theorem}\label{thm.4.1}
Assume that the condition {\rm (\ref{eq:4.12}) } is satisfied,
%$J > (1 + 2 r_{\rm max}) J'$,
then,  for any 
$\b > \b_0= 1.9 \,  (1+\r)[J_{AB}-\r (J_{AW} - J_{BW})]^{-1}$, 
the following serie, defining the wall-medium surface tension,
is absolutely convergent
\begin{equation}
\st_{AW} =  %&= &
r J_{AW} 
- \frac{1}{\b a}
\sum_{b \in W_a}  \sum_{C \ni b} 
\frac{\tf (C)}{|C \cap \wall|} 
\label{eq:4.13}
%\\
%&=&
%r J_{AW}  
%- \frac{1}{\b a}
%\sum_{C \subset \L_a  } 
%\tf (C)  
%- \frac{1}{\b a}
%\biindice{\dis \sum}{C \cap \L_a \not= \emptyset}{C \cap \L_a^c \not= \emptyset}  
%\tf (C) \frac{|C \cap W_a|}{|C \cap \wall|}  
\end{equation}
\end{theorem}

\proof
The proof of formula (\ref{eq:4.13}) as well as that of the absolute convergence
of the series  can be established following  \cite{GMM} (Chapter 4)
in which the low temperature contours of the Ising model were considered
in the role played here by the excitations (see also \cite{D,KP,Simon}).
\newline
The first ingredient is the following lower bound on  the energy:
\begin{equation}
E(\d) \geq (1+\r)^{-1} [J_{AB}-\r (J_{AW}-J_{BW})] |\d|
\label{eq:4.14}
\end{equation}
This bound follows from expression 
$$
E(\d) 
=
 \left[  J_{AB}-
\frac{J_{AB}+J_{AW}-J_{BW}}{1+\frac{|\d \setminus (\d \cap W)| }{|\d \cap w|} } 
\right]
\pes |\d| 
$$
obtained from (\ref{eq:4.6}) and the inequality
$ (|\d \cap W| / |\d \setminus (\d \cap W)| ) \leq \r$
that holds true for any excitation $\d$: this 
inequality is a consequence of easy geometrical arguments used with condition
(\ref{eq:4.12}).
% and with
% the  inequality 
% $(\sum_k a_k) / (\sum_k b_k) \leq \sup_k (a_k/b_k)$
% valid for positive numbers $a_k, b_k$.
\newline
Inequality (\ref{eq:4.14}) together with the fact that 
the number of polygons (or of excitations $\d$) of length $\ell$
passing to a given point is less then $3^\ell$
ensures in particular
the convergence of the series 
$
\sum_{\d \ni b} e^{-\b E(\d)}
$,
for all bond $b$
as soon as $\b$ equals some $\b'_0$.
\newline
The convergence of the cluster expansion  needs furthermore
the existence of a positive real-valued function
$\mu (\d)$  such that
\begin{equation}
	e^{-\b E(\d)} \mu (\d)^{-1} \exp 
	\bigg\{
	\sum_{\d'\, i \;  \d}\mu (\d)
	\bigg\}
	\leq e^{-\a} < 1
	\label{eq:4.15}
\end{equation}
where the sum runs over excitations $\d'$ incompatible with $\d$:
this relation is denoted by $\d'\, i \;  \d$ and means 
that $\d'$ do not intersect $\d$.
Taking in addition with the above remark on the entropy of excitations, 
that the lengths of excitations are even with minimal value
$|\d_{\rm min}|=4$,
that 
$\sum_{\d' i \d}\mu (\d) \leq |\d | \sum_{\d' \ni b}\mu (\d')$,
and choosing 
$\mu(\d) = (3e^{t})^{-|\d|}$, inequality (\ref{eq:4.15}) will be 
satisfied whenever
\begin{displaymath}
	\b \, (1+\r)^{-1} [J_{AB}-\r (J_{AW}-J_{BW})]
	>
	\log 3 + t+ \frac{e^{-4t}}{1-e^{-2t}}
\end{displaymath}
The value $t_{0}\simeq .61$ that minimizes the function 
\begin{math}
	t+[e^{-4t}/(1-e^{-2t})]
\end{math}
provides the corresponding $\b_{0}$ given in the theorem.
The expression \eqref{eq:4.13} then follows from \eqref{eq:4.2} and 
\eqref{eq:4.9} by setting $N=pa$ and letting $p\to \infty$.
\endproof

The other cases where the ground state is not the wall  $W$ 
will be discussed elsewhere \cite{DMR}. They  lead 
in particular to the  
Cassie's
law \cite{C}.

\section{Comparison of different geometries} 
\setcounter{equation}{0}
\setcounter{dessin}{0}

We shall now restrict to some specific walls $W$.
Namely, we assume $\bar{h}_i = 0$ for $i=0,\dots, c-1$ and 
$\bar{h}_i = b$ for $i=c,\dots, a-1$,
($1\leq c \leq a-1$), see Fig.~\ref{F5.3}. 

We will denote 
by
$	\D \tau (r,c)=\st_{AW}-\st_{BW}$ 
the difference between the  surface tensions
corresponding to the roughness $r$ 
and the  parameter~$c$.
The roughness has value $r= 1+ 2b/a$ 
and 
it is  independent of $c$.

Let us stress that Theorem~\ref{thm.4.1} implies Wenzel's  law 
at low enough temperature
$$
\D \st (r, c) = r (J_{AW}-J_{BW}) + {\rm corrections}
$$

The next theorem concerns the corrections to Wenzel law.
It gives a  comparison between different geometries 
with the same roughness by varying the parameter~$c$.

\begin{theorem}\label{thm.5.1}
	\setcounter{equation}{0}
Assume that $J=J_{AB}>0$ and $J'=J_{AW} - J_{BW}>0$
satisfy 
$J-(2b+1) J' \equiv 2(b+1) K_{1} >0$,
%$\beta > \beta _1= cte_{1} \, (b+1)[J_{AB}-(2b+1) (J_{AW} - 
%J_{BW})]^{-1}+ cte_{2}$ 
then, for $\b$ large enough, 
$\beta M >  (b+1)(1.9 (a+c)+ .56)$
where 
$M = \min \{ (b+1) J', 2(b+1) K_{1}, |J' -(b+1) K_{1}|\}$:
\begin{description}
	\item[{\rm a)}]
if $2 \leq c  \leq a-1$,
\[
	\D \tau (r,1)  < \D \tau (r,c) 
\]
	\item[{\rm b)}]  
	if $1\leq c  \leq c_{0} -1$,
\[ \D \tau (r,c)  < \D \tau (r,c+1) \]
	\item[{\rm c)}]
if $c_{0} \leq c  \leq a-2$,
\[
	\D \tau (r,c)  > \D \tau (r,c+1)
\]
\end{description}
where  when $a$ is odd $c_{0}= (a+5)/2$
and when  $a$ is even 
 $c_{0}= a/2 +2$ if $J' <(b+1) K_{1}$ 
 and $c_{0}= a/2 +3$ if $J' > (b+1) K_{1}$.
\end{theorem}

This result is represented graphically in Fig.~\ref{F5.1}
\begin{center}
\setlength{\unitlength}{6.5mm}
\begin{picture}(15,12)
    \put(0,0){
\begin{picture}(15,12)
    %exitation d0
\thinlines
 \drawline(0,0)(15,0)
 \drawline(15,0)(15,12)
\drawline(15,12)(0,12)
\drawline(0,12)(0,0)

\drawline(1,0)(1,0.1)
\drawline(2,0)(2,0.1)
\drawline(3,0)(3,0.1)
\drawline(4,0)(4,0.1)
\drawline(5,0)(5,0.1)
\drawline(6,0)(6,0.1)
\drawline(7,0)(7,0.1)
\drawline(8,0)(8,0.1)
\drawline(9,0)(9,0.1)
\drawline(10,0)(10,0.1)
\drawline(11,0)(11,0.1)
\drawline(12,0)(12,0.1)
\drawline(13,0)(13,0.1)
\drawline(14,0)(14,0.1)

\put(1.8,-.5){$_{2}$}
\put(3.8,-.5){$_{4}$}
\put(5.8,-.5){$_{6}$}
\put(7.8,-.5){$_{8}$}
\put(9.7,-.5){$_{10}$}
\put(11.7,-.5){$_{12}$}
\put(13.7,-.5){$_{14}$}

\put(15.5,0){$c$}
\put(1,10.5){$\Delta \tau (r,c)$}
%les points
\put(1,0.73672){\circle*{.2}}
\put(2,2.8806){\circle*{.2}}
\put(3,4.67133){\circle*{.2}}
\put(4,6.16707){\circle*{.2}}
\put(5,7.41641){\circle*{.2}}
\put(6,8.45995){\circle*{.2}}
\put(7,9.33159){\circle*{.2}}
\put(8,10.0596){\circle*{.2}}
\put(9,10.4295){\circle*{.2}}
\put(10,10.4533){\circle*{.2}}
\put(11,10.3699){\circle*{.2}}
\put(12,10.1767){\circle*{.2}}
\put(13,9.86729){\circle*{.2}}
\put(14,9.43174){\circle*{.2}}

\end{picture}
}
\end{picture}
\end{center}
\begin{dessin}
\label{F5.1} 
\figcaption{Plot of the wall tensions difference $\D \st (r,c)$ as 
function of the parameter $c$.}
\end{dessin}
It means that for a given roughness $r$, there is an optimum 
in the wall tension at $c=c_{0}$. 
That is to say that for 
$c=c_{0}$, the associated contact angle $\t$ for the sessile drop will 
be
minimum for $\t < \pi/2$ and maximum for $\t> \pi /2$.
These results also confirm the data obtained by numerical simulations 
in \cite{TUBD}.
It was indeed observed that $\D \st (r)$ for  $2d$ random 
substrates with a fixed roughness $r$ remains between the ``single''
protusion and ``single'' hole case as represented in Fig.~\ref{F5.2}.

\begin{center}
	\epsfig{file=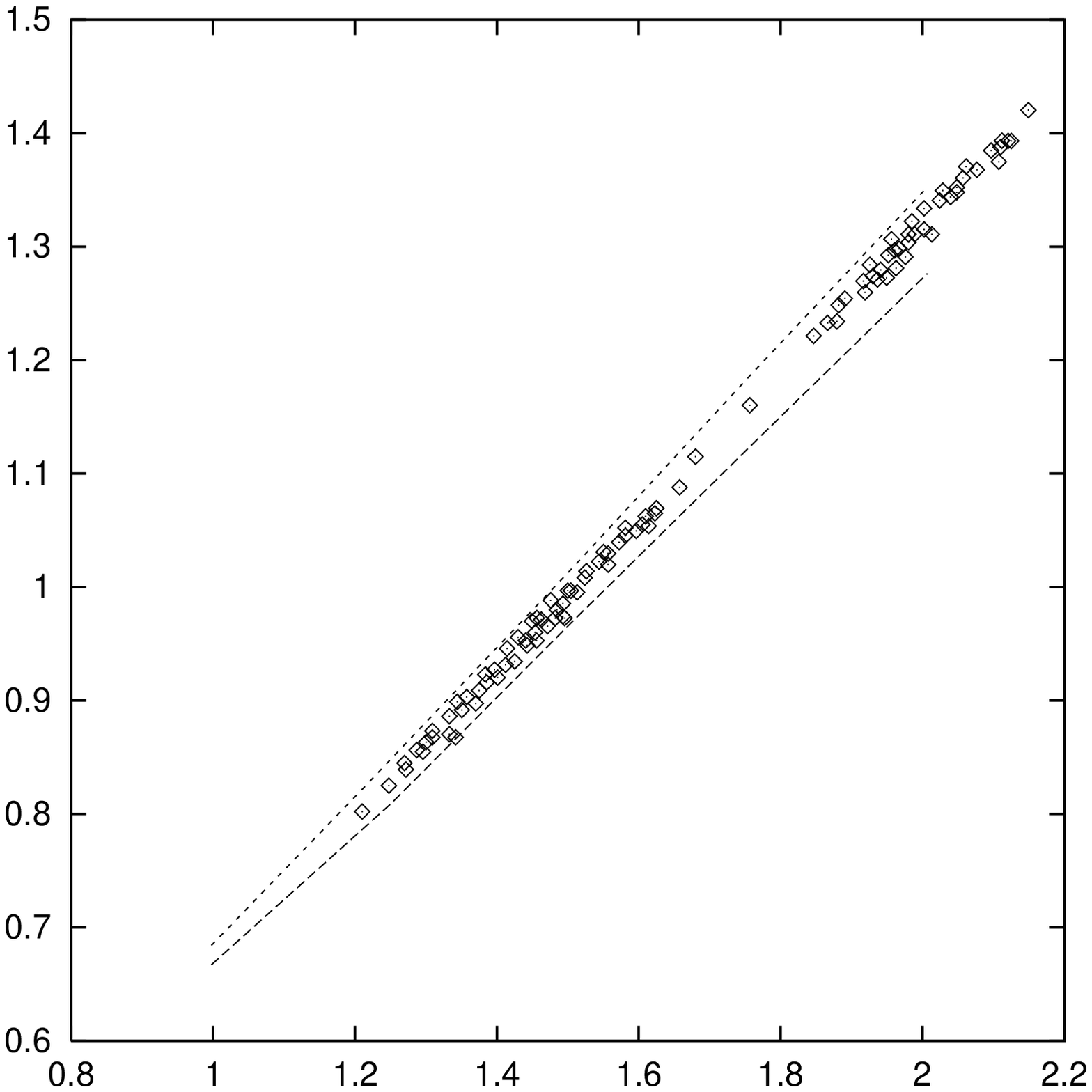,height=8cm}
	\\[1ex]
	\epsfig{file=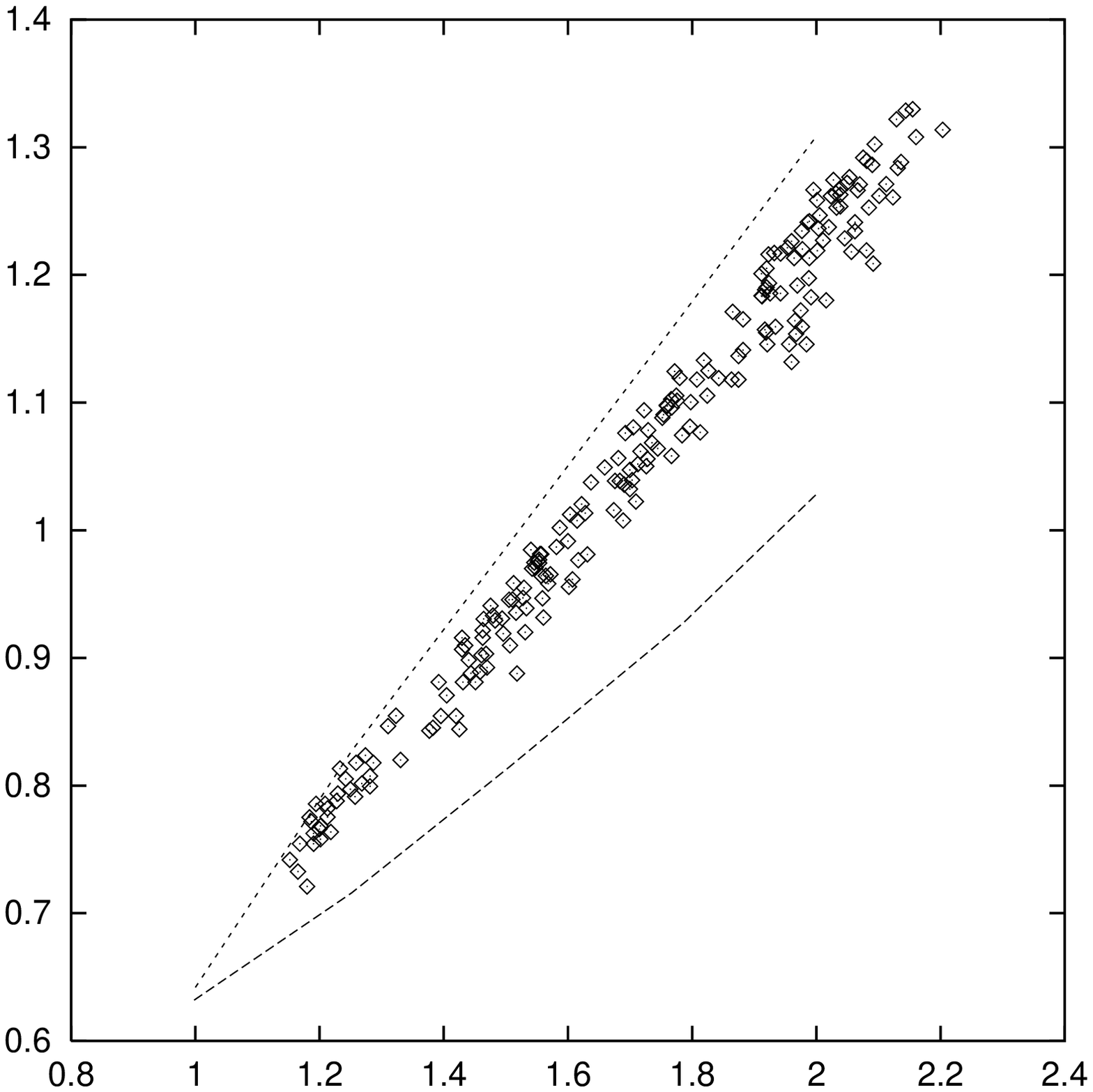,height=8cm}
\end{center}

\begin{dessin}
\label{F5.2} 
\figcaption{Plot of  $\D \st (r)$ as 
function of the roughness $r$ at two different temperatures, 
for the single hole $c=1$ (lower dashed lines), 
the single protusion $c=a-1$ (upper dashed lines) and the random geometry 
(diamond),
\cite{TUBD}.}
\end{dessin}
On the basis of Theorem~\ref{thm.5.1}, we have in fact
$$
\D \st (r,c=1) \leq \D \st (r,c) \leq \D \st (r, c_{0})
$$
Since on the other hand  we have 
$$
\D \st (r,c=a-1) \simeq \D \st (r,c=c_{0})
$$
we can understand that indeed single protusions and single holes
will be good approximations for the upper and lower limits of the wall tension
as already indicated in \cite{TUBD}.

Let us also point out here that the numerical simulations 
seem to indicate that the first order correction to 
Wenzel is already enough to describe the wall tensions up to the 
half of the $2d$ Ising critical temperature.

That this optimal geometry also holds for more general systems 
remains up to now an interesting open question.

Before proving Theorem~\ref{thm.5.1}, we give the following

\begin{lemma}
	\label{lem.5.1}
Assume that  
$cJ-(c+ 2b)J'\equiv 2(c+b)K_{c} >0$,
then, for $\b \geq  2(1.9+\a/4)(b+c)[cJ-(c+2bJ')]^{-1}$, $\a>0$ we have, 
\begin{equation}
\sum_{\substack{C\ni (0,0)\\ C\ni (0,c)}} \tf (C)
=
[1+\ve(c)]\,  e^{-\b [cJ - (c+2b) J']}
% =
% (1+\ve_{\b}) e^{-2\b (c+b) K}
\label{eq:5.1}
\end{equation}
where
\begin{align*}
&|\ve(c)|\leq
	e^{2\nu(c+b)+\eta} 
	\Big[ 
	e^{-2(\b K_{c} -\nu)}  + \frac{2}{1-e^{-\b(J+J')}}
	\big( 
	e^{-\b (J+J')} + 
	\frac{e^{-2(\b J'+\nu) }}{1-e^{-2(\b J'+\nu)}} 
	\big)
	\Big]
\\
&\nu= 1.9+\a/4, 
\quad
\eta 
= \log \frac{e^{-4t_{0}} }{1-e^{-t_{0}}}\frac{\, e^{-\a} }{1-e^{-\a}} 
=\log \frac{0.2 \, e^{-\a} }{1-e^{-\a}} 
\end{align*}
\end{lemma}

\begin{proof}
We first observe
%Notice 
that
all excitations 
%$\d$  
satisfy
 $\max_{\d} (|\d \cap W| / |\d \setminus (\d \cap W)| ) \leq (c+2b)/c$,
 i.e., 
	$\r = 1+ 2b/c$,
	so that inequality \eqref{eq:4.14} reads
	\begin{equation}
E(\d) \geq \frac{1}{2(b+c)} [cJ-(c+2b) J'] |\d|= K_{c}  |\d|
\label{eq:5.2}
\end{equation}
\noindent
Under the condition,  
$\beta K_{c} \geq 1.9 +\a /4$, $\a >0$,
the cluster expansion converges
and moreover
\begin{equation}
	\sum_{\substack{|C|\geq m\\C \cap b\not= \emptyset}} |\tf (C)|
	\leq  
	%\frac{0.2 \, 	e^{-\a} }{1-e^{-\a}}   
\, 	\exp\{(-\b K_{c} + \nu)\,  m + \eta\}
	\label{eq:5.3}
\end{equation}
where for a cluster $C=\{\d_{1},\dots,\d_{n}\}$ we use 
$|C|= |\d_{1}|+ \cdots +|\d_{n}|$ to denote its length.
 To prove  \eqref{eq:5.3}, we write for a cluster 
 $C=\{\d_{1},\dots,\d_{n}\}$ of length at least m,
 $|\tf(C)|\leq (1/n!) e^{-(\b K_{c} - \b_{1}K_{c})m} 
 a(\d_{1},\dots,\d_{n})\prod_{i=1}^{m} e^{-\b_{1} E(\d_{i})}$.
 Then we use that condition \eqref{eq:4.15} is satisfied if for any 
 $\d$,
 $\b K_{c} \geq \log 3 +t  + \frac{e^{-4t}}{1-e^{-2t}} + \a / |\d|$,
 i.e.\
 for $\b K_{c} \geq 1.9 + \a / 4$ by choosing $\mu$ and  $t$ as in the proof of 
 Theorem~\ref{thm.4.1}.
 Under \eqref{eq:4.14},  one knows, c.f.\ \cite{GMM}, that
 $\sum_{C\ni \d}\tf (C) \leq \mu (\d) e^{-\a}/(1-e^{-\a})$.
 	We use finally $\sum_{\d \ni b}\mu (\d)\leq e^{-4t}/(1-e^{-2t})=.2$ when 
	$t=.61$ to get \eqref{eq:5.3}.

\noindent
We let
    $\d_{0}$ 
be the excitation corresponding to the interface 
    $\G_{0}$, 
    $\d_{0} = (\G_{0}\cup \wall) \setminus (\G_{0}\cup 
    \wall)$,
where 
    $\G_{0}$
is given by the height
    $h_{i}=b$
for 
   $i=0,1,\dots,c-1 $
and 
  $	h_{i}=\bar{h}_{i}$
otherwise.
That is
    $\d_{0}$
is the boundary of the rectangle
	$	R = \{ (x,y) \in \reel^{2} \, : \, 0\leq x \leq c, \,  0,\leq y \leq b \}$,
see	Fig.~\ref{F5.3}.
Its energy is: 
	$	E(\d_{0}) = c J - (c+2b) J'$.
	Denote by
$W(b_{1},b_{2})$ the part of the wall between the points 
$B_{1}=(0,b_{1})$ and $B_{2}=(c,b_{2})$.
Then,
\begin{equation}
	\sum_{C: C\cap W \supset  W(b,b) } 
	\tf (C)
	=
	 e^{-\b [cJ - (c+2b) J']} %e^{-\b E(\d_{0})}
	+
	\sum_{\substack{C:  C\cap W  \supset  W(b,b)\\ |C|\geq 2(c+b+1)} } 
	 \tf (C)
	\label{eq:5.4}
\end{equation}
Indeed the first term of the R.H.S. of \eqref{eq:5.4}
corresponds to the excitation $	\d_{0}$ and the second terms run over 
the other clusters containing $W(b,b)$ the length of them being at least
$|\d_{0}|+2=2(c+b+1)$.
By \eqref{eq:5.3}, this term is bounded as follows:
	\begin{equation}
	\sum_{\substack{C:  C\cap W  \supset  W(b,b)\\ |C|\geq 2(c+b+1)} } 
	 |\tf (C)|
% 	 \leq
% 	 \exp\{-2(\b K- 1.9 - \a/4)(c+b+1)\}
% 	 % e^{2(1.9+\frac{\a}{4})(c+b+1)} 
% 	  \sum_{\d \ni b}\mu (\d) \frac{e^{-\a}}{1-e^{-\a}}
	  %\\
	  %\leq
% 	  \exp\{
% 	  -[\b 
% 	  \frac{cJ-(c+2b)J'}{c+b}
% 	  -(3.8+\a/2 )] (c+b+1)
% 	  \}
% 	   \frac{e^{-\a}}{1-e^{-\a}} \frac{e^{-4t}}{1-e^{-t}}
% 	   \\
	   \leq
	   %\frac{0.2 \, e^{-\a}}{1-e^{-\a}} 
	   \, \exp \big\{
	   -2 (c+b+1)(\b K_{c}-\nu) + \eta \big\}
	   %\frac{e^{-\a}}{1-e^{-\a}} 
	   %\frac{e^{-4t}}{1-e^{-t}}
		\label{eq:5.5}
	\end{equation}

	Next we observe that for all excitations whose intersection with
	the wall is $W(b_{1},b_{2})$ satisfy 
	 $\max_{\d} (|\d \cap W| / |\d \setminus (\d \cap W)| ) \leq 
	 (c+b_{1}+b_{2})/(c+|b_{1}-b_{2}|$.
Hence the bound \eqref{eq:5.2}  is improved as follows:
	\begin{equation}
E(\d) \geq \frac{ (c+|b_{1}-b_{2}|)J-(c+b_{1}+b_{2}) J'}{(2c+b_{1}+b_{2}+|b_{1}-b_{2}|)}
|\d| 
\label{eq:5.6}
\end{equation}
when $\d \cap W = W(b_{1},b_{2})$.
All the associated clusters have length 
$|C|\geq 2c+b_{1}+b_{2}+|b_{1}-b_{2}|$.
Therefore
\begin{multline}
	\sum_{C:C \cap W = W(b_{1},b_{2}) } 
	| \tf (C)|
	 \leq
	 %\frac{0.2 \, e^{-\a}}{1-e^{-\a}} 
	 \exp \{-\b [(c+|b_{1}-b_{2}|)J-(c+b_{1}+b_{2})J'] \}
	 \\
	 \times
	  \exp \{ (2c+b_{1}+b_{2}+|b_{1}-b_{2}|)\nu + \eta\} 
	  % \frac{e^{-\a}}{1-e^{-\a}} \frac{e^{-t}}{1-e^{-4t}}
		\label{eq:5.7}
	\end{multline}
	Thus,
		\begin{multline}
		\sum_{\substack{ b_{1}+b_{2}\leq 2b-1 \\ b_{1} \geq 1,b_{2}\geq 1}}
		\sum_{C:C \cap W = W(b_{1},b_{2}) }
		 |\tf (C)|
		 \leq
		 %2\times 1.023 \times  
		 %\frac{2 }{1-e^{-\b (J+J')}} 
		 2 \, \big[1-e^{-\b (J+J')} \big]^{-1} 
		 \\
		 \times \Big(
		 \exp \left\{ -\b (c+1)J +\b (c+2b-1)J' +2\nu (c+b) + \eta \right\}
		 \\
		 +
		 %\frac{1 }{1-e^{-2\b (J'+\nu)}}  
		 \big[1-  e^{-2 (\b J'+\nu)}\big]^{-1} 
		 \exp \left\{- \b c J +  (c+2b-2)J' +2\nu (c+b-1)+\eta \right\}
		 \Big)
		\label{eq:5.8}
	\end{multline}
	The first term inside the parenthesis comes from the summation over 
	$1\leq b_{2}\leq b_{1}-1$, $b_{1}=b$ and the second term from
	the summation over 	$1\leq b_{2}\leq b_{1}$, $1\leq b_{1}\leq b-1$.
% 	This is done by suming  geometric series giving additional terms of 
% 	type $(1-e^{-x})^{-1}$ which are bounded by $1.023$ because 
% 	$x\geq 3.8$
% 	in all the considered cases.
	
	Using that
	$$
\sum_{\substack{C\ni (0,0)\\ C\ni (0,c)}} \tf (C)
= 
\sum_{C: C \cap W \supset  W(b,b) } 
	\tf (C)
	+
\sum_{\substack{ b_{1}+b_{2}\leq 2b-1 \\ b_{1} \geq 1,b_{2}\geq 1}}
		\sum_{C:C \cap W = W(b_{1},b_{2}) }
 \tf (C)
	$$
	the proof follows from (\ref{eq:5.4},\ref{eq:5.5},\ref{eq:5.8}).
\end{proof}

\begin{proofof}{Theorem \ref{thm.5.1}}

The lower bounds (\ref{eq:5.2},\ref{eq:5.6}) on the energy 
can be improved for some excitations.
Let
$\L(i,j) = \{ x=(x,  y) \in \Reel^2 : i \leq x \leq j \}$
denote the infinite cylinder between the vertical lines
$x=i$ and $x=j$.
For the excitations included in the strips
$\L(1-a,c-1)$ and $\L(1,a+c-1)$,
one has
$\max_{\d} (|\d \cap W| / |\d \setminus (\d \cap W)| ) \leq 1$,
and thus by arguing as in the proof of \eqref{eq:4.14}
\begin{equation}
E(\d) \geq \frac{1}{2} (J-J') |\d|	
	\label{eq:5.9}
\end{equation}
The associated clusters satisfy thus:
\begin{equation}
	\sum_{\substack{|C|\geq m\\C \ni b}} |\tf (C)|
	\leq   \exp \Big\{-\frac{\b}{2}(J-J') \, m + \nu\,  m  + \eta\Big\}
	\label{eq:5.10}
\end{equation}
For the excitations included in the strips
$\L(1,c-1)$ and $\L(c,a)$,
one has
$|\d \cap W| \leq |\d| / 2 -1$.
Therefore
\begin{equation}
E(\d) =  J |\d|	 -(J +J')|\d \cap W | \geq \frac{1}{2} (J-J') |\d|+J+J'	
	\label{eq:5.11}
\end{equation}
and the associated clusters satisfy:
\begin{equation}
	\sum_{\substack{|C|\geq m\\C \ni b}} |\tf (C)|
	\leq   \exp \Big\{- \frac{\b}{2}(J-J') \, m - \b (J+J') + \nu\,  m  + 
	\eta \Big\}
	\label{eq:5.12}
\end{equation}
	We let
    $\d_{1}$ 
be the excitation corresponding to the interface 
    $\G_{1}$, 
    $\d_{1} = (\G_{1}\cup \wall) \setminus (\G_{1}\cup \wall)$,
where 
    $\G_{1}$
is given by the height
	$h_{i}=b+1$ for   $i=c-1, \dots,a-1 $,
and 
  $	h_{i}=\bar{h}_{i}$
otherwise (Fig.~\ref{F5.3}).
Its energy is 
	$	E(\d_{1}) = (a-c+b+3) J - (a-c+b+1) J'$ and $|\d_{1}|=2(a+b+c+2)$.
\newline
	We let 
	$\d^{(k)}(x,y)$
	denote the excitations of width $k$, height $1$ (and length 
	$2(k+1)$) whose intersection with the wall is the segment 
	$[(x,y),(x+k,y)]$ (of length $k$).
	Their energy when $y=b$ or when $y=0$ and they do not intersect the 
	vertical part of the wall are:
	$E^{(k)}= k(J-J') + 2J$.
	
	%\bigskip
\begin{center}
\setlength{\unitlength}{8.5mm}
\begin{picture}(14,7)
    \put(0,0){
\begin{picture}(0,0)
    %excitation d0
\thinlines
\drawline(-0.5,-1)(-0.5,5.5)
\drawline(-0.5,5.5)(14.5,5.5)
\drawline(14.5,5.5)(14.5,-1)
\drawline(-0.5,-1)(14.5,-1)
\thicklines
\put(0,4){\line(1,0){1}}
\put(1,0){\line(0,1){4}}
\put(1,0){\line(1,0){3.5}}
\put(4.5,0){\line(0,1){4}}
\put(4.5,4){\line(1,0){2.5}}
\thinlines
\put(7.1,0.1){\line(0,1){3.9}}
\put(7.1,0.1){\line(1,0){3.3}}
\put(10.4,0.1){\line(0,1){3.9}}
\put(7.1,4){\line(1,0){3.3}}
\put(8.7,4.2){$\delta_{0}$}

%points
\put(6.5,-.5){\small $(0,0)$}
\put(10,-.5){\small $(c,0)$}
\put(12.5,-.5){\small $(a,0)$}

\drawline(4.5,1.4)(4.7,1.3)
\drawline(4.5,1.4)(4.7,1.5)
\drawline(4.5,1.4)(4.9,1.4)      
  \drawline(6.8,1.3)(7,1.4)   
  \drawline(6.8,1.5)(7,1.4) 
  \drawline(6.6,1.4)(7,1.4) 
  \put(5.2,1.3){$a-c$}

\put(2.7,-.5){$c$} 
\drawline(1,-.4)(1.5,-.4)
\drawline(1,-.4)(1.2,-.3)
\drawline(1,-.4)(1.2,-.5)
\drawline(4,-.4)(4.5,-.4)      
  \drawline(4.3,-.3)(4.5,-.4)   
  \drawline(4.3,-.5)(4.5,-.4)    
\put(0.4,2){$b$}
\drawline(0.5,3.5)(0.5,3.99)   
   \drawline(0.4,3.8)(0.5,3.99)   
   \drawline(0.6,3.8)(0.5,3.99)
\drawline(0.5,0)(0.5,0.5)
\drawline(0.5,0)(0.4,0.2)
\drawline(0.5,0)(0.6,0.2)

%excitation d1
\thicklines
 \put(7,0){\line(0,1){4}}
 \put(7,0){\line(1,0){3.5}}
 \put(10.5,0){\line(0,1){4}}
 \put(10.5,4){\line(1,0){2.5}}
 \drawline(13,4)(13,0)
 \drawline(13,0)(14,0)
 
 \thinlines
 \drawline(4,0.1)(4,4.6)
 \drawline(4,4.6)(7,4.6)
 \drawline(7,4.6)(7,4.1)
  \drawline(7,4.1)(4.4,4.1)
  \drawline(4.4,4.1)(4.4,0.1)
  \drawline(4.4,0.1)(4,0.1)
  \put(3.5,4.5){$\delta_{1}$}
  %excitation dk
  \thinlines
 \drawline(11,4.6)(12.5,4.6)
 \drawline(12.5,4.6)(12.5,4.1)
  \drawline(12.5,4.1)(11,4.1)
   \drawline(11,4.1)(11,4.6)
   \put(12.7,4.5){$\delta^{(k)}$}
 \end{picture}
}
\end{picture}
\end{center}
\begin{dessin}
\label{F5.3} 
\figcaption{The excitations $\d_{0}$, $\d^{(k)}$, and 
$\d_1$ translated by $-a$.}
\end{dessin}

	Then we the decompose the sum involved in \eqref{eq:4.13} 
as follows:
\begin{equation}
	\sum_{C\cap \L(0,a)\not= \emptyset}  \tf (C)
	=S_{1}(c) + S_{2}(c)+S_{3}(c)
	\label{eq:5.13}
\end{equation}
where
\begin{align}
	S_{1}(c)&= 
	\sum_{C\subset \L(1,c-1) }  \tf (C)
	+\sum_{C\subset \L(c,a) }  \tf (C)
	\nonumber
	\\
	S_{2}(c)&=
	\sum_{\substack{C\subset \L(1-a,c-1)\\ C\cap (0,0)\not= \emptyset}}  \tf (C)
	+\sum_{\substack{C\subset \L(1,a+c-1)\\C\cap (0,c)\not= \emptyset}}  \tf (C)
	\nonumber
	\\
	S_{3}(c)&=%e^{-\b[c(J-J')-2bJ']}
	\sum_{\substack{C\cap (0,0) \not= \emptyset\\ C \cap (c,0) \not= \emptyset}}  \tf (C)
	\nonumber
\end{align}
Let us compare the differences $S_{i}(c)- S_{i}(c+1)$, $i=1,2,3$. 
When $a$ is even, we have
\begin{equation}
		S_{1}(c)- S_{1}(c+1)= 
%\left\{
	\begin{cases}
		e^{-\b c(J-J')-2\b J} + R_{1}(c)
	&{\rm if} \ c-2 \leq a-c-2  
	\\
	 - e^{-\b (a-c+1)(J-J')-2\b J}
	+ R_{1}'(c)
	& {\rm if} \ c-2 \geq a-c  
 \end{cases}
% \right.
 \label{eq:5.14}
\end{equation}
where
\begin{equation}
%	\left\{
	\begin{align}
	&|R_{1}(c)|\leq 
	4 %\sideset{}{^{(3)}} 	
	\sum_{\substack{C\subset \L(c,a)\\ |C|\geq 2(c+2)} } 
	| \tf (C)|
	\leq 4 e^{ - \b (c+2) (J-J') -\b(J+J')+ 2\, (c+2) \nu +\eta }
	\label{eq:5.15}
	\\
	&|R_{1}'(c)|\leq 
	4 %\sideset{}{^{(3)}} 	
	\sum_{\substack{C\subset \L (1,c) \\ |C|\geq 2(a-c+3)} } 
	| \tf (C)|
	\leq 4 e^{ - \b (a-c+3) (J-J') -\b(J+J') + 2\, (a-c+3)\nu  +\eta}
	\label{eq:5.16}
 \end{align}
% \right.
\end{equation}
Indeed, there is a one--to--one correspondence
between the  clusters $C$ of base of size $|C\cap W|=k$
occurring in $S_{1}(c)$ and $S_{1}(c+1)$ till $k$ reach some value.
This value, when $c-2 \leq a-c-2 $ is precisely $c$, because in that 
case there are clusters (of base of size $c$) which belong to 
$\L(c,a)$ bot neither to $\L(1,c-1)$ nor to $\L(c+1,a)$.
There is precisely one excitation $\d^{(c)}$ of base of size $c$ (and length $2(c+1)$)
which belong to 
$\L(c,a)$ bot neither to $\L(1,c-1)$ nor to $\L(c+1,a)$.
Its energy is $E^{(c)}=c(J-J')+2J$ and gives the corresponding term in \eqref{eq:5.14}.
The other clusters have length $|C|\geq 2(c+2)$. This gives the first bound 
on the reminder $R_{1}(c)$.
The second bound in \eqref{eq:5.15} follows from \eqref{eq:5.12}.
When $c-2 \geq a-c $ the argument works in the opposite direction.
The value $k$ is $a-c+1$ and there is a corresponding $\d^{(k)}$ (of 
length $2(a-c+2)$ which 
belong to $\L(1,c)$ but nor to $\L(1,c-1)$ nor to $\L(c,c-a)$.
Its energy is $(a-c+1)(J-J')+2J$ and provides the corresponding term in
 \eqref{eq:5.14}.
The other clusters have length $|C|\geq 2(a-c+3)$. This gives the bound 
on the reminder $R_{1}'(c)$, the second inequality in \eqref{eq:5.16} 
following  from \eqref{eq:5.12}.
\newline
When $a$ is odd:
\begin{equation}
	S_{1}(c)- S_{1}(c+1) =
	\begin{cases}
	e^{ -\b c(J-J')-2\b J}+	R_{1}(c)
	&{\rm if} \ c-2 \leq a-c-3  
	\\
	0 &{\rm if} \ c-2 = a-c -1  
	\\
	- e^{-\b (a-c+1)(J-J')-2\b J}+  R_{1}'(c)	
	&{\rm if} \ c-2 \geq a-c+1  
 \end{cases}
 \label{eq:5.17}
\end{equation}
where  $0$ must be understood as 
$\sum_{C \geq M } \tf (C)$ with $M$ as large as we wish.
Indeed
 the same reasoning as for $a$ even applies.
In addition there is the particular case  $c-2 = a-c -1 $
where the width of the cylinder $\L(1,c-1)$ equals the one of $\L(c-a-1)$
and the width of $\L(c-a)$ equals the width of $\L(1,c)$.
\newline
For $S_{2}(c)$, we have:
\begin{multline}
|S_{2}(c)- S_{2}(c+1)| =|R_{2}(c)|\leq
2%\sideset{}{^{(4)}}	
	\sum_{\substack{C \cap(0,c)\not= \emptyset \\ |C|\geq 2(c+1)} } 
	| \tf (C)|
	+ 
	2%\sideset{}{^{(4)}}	
	\sum_{\substack{C \cap(0,c)\not= \emptyset \\ |C|\geq 2(a-c+b+2)} } 
	| \tf (C)|
	\\
	\leq
	2 e^{ - \b (c+1) (J-J')+ 2 \, (c+1) \nu + \eta }
    +
    2 e^{ - \b (a-c+b+2) (J-J') -\b(J+J') + 2\, (a-c+b+2) \nu + \eta }
	\label{eq:5.18}
\end{multline}
Indeed the  clusters of minimal energy containing $(0,c)$ and
for which the
correspondence is not one--to--one are the
excitations $\d^{(c)}(0,0)$ of length $2(c+1)$ and the excitation
$\d_{1}$ of length $2(a-c+b+2)$.
All other clusters have length greater or equal than either
$|\d^{(c)}(0,0)|+2$ or than $|\d_{1}|+2$.
To bound the first sum we used \eqref{eq:5.10} 
and to bound the second sum we used \eqref{eq:5.12}.

Finally, by Lemma~\ref{lem.5.1}
\begin{equation}
 S_{3}(c)- S_{3}(c+1) 
 %\gtreqqless
=(1+ \ve') e^{-\b [cJ - (c+2b) J']}
	%\pm 2 %\sideset{}{^{(2)}}	
%+R_{3}(c)
%\sum_{\substack{C \cap(0,0)\not= \emptyset \\ C \cap(c,0)\not= \emptyset\\|C|\geq 2(c+2) } } 
	%| \tf (C)|
	\label{eq:5.19}
\end{equation}
where 
\begin{equation}
	|\ve'|
%	\leq
%	e^{ - \b (c+1) (J-J') +\b {(c+1)^{-1}} (J-J') 
%	 + \b \frac{c+2}{c+1} J' + 3.8\, (c+2) }
%	 \\
	 \leq
	| \ve(c) |+ e^{ - \b  (J-J') } (1+ |\ve (c+1)|)
	\label{eq:5.20}
\end{equation}
% The first term of the R.H.S. of \eqref{eq:5.20} corresponds to the excitation
% $\d_{0}(c)$ of length $2(c+b+1)$.
% If $b=1$ the other clusters for which the correspondence is not 
% one--to--one have length $2(c+2)$.
% This gives the first bound on the reminder while the second inequality in
% \eqref{eq:5.20} follows from \eqref{eq:5.3}.
% 

Assume now that 
$2c \geq a +2$ if $a$ is even or
$2c \geq a +3$ if $a$ is odd, then from 
\eqref{eq:5.14}--\eqref{eq:5.20}
\begin{multline} 
	\D\tau(c+1)-\D\tau(c)
	=\sum_{i=1}^{3} S_{i}(c)-S_{i}(c+1)
% 	\\
% 	%\gtreqqless
% 	\lesseqqgtr
% 	 -e^{-\b (a-c+1)(J-J') - 2\b J'} + e^{-\b c'(J-J')+2\b J'}	
% 	 \pm (|R_{1}'(c)|+|R_{2}(c)|+|R_{3}(c)|)
% 	 \\
% 	=
% 	 -(1- \ve_{\b}) e^{-\b (a-c+1)(J-J') - 2\b J'} 
% 	 + (1+\ve_{\b}') e^{-\b c'(J-J')+2\b J'}
% 	 \label{eq:5.15}
% \end{multline}
% so that
% \begin{align}
% \D\tau(c+1)-\D\tau(c)
	\\ = %\leq
	 -(1+ \ve_{1}) e^{-\b (a-c+1)(J-J') - 2\b  J} 
	 + (1+\ve_{2}) e^{-\b c(J-J')+2\b b J'}
% 	 \\
% 	 \D\tau(c+1)-\D\tau(c)
% 	 &\geq
% 	 -(1+ \ve_{1}) e^{-\b (a-c+1)(J-J') - 2\b J'} 
% 	 + (1-\ve_{2}) e^{-\b c'(J-J')+2\b J'}
	\label{eq:5.21}
\end{multline}
where
\begin{align}
	&|\ve_{1}|\leq
	  4 e^{-\b (J -J')  + 2 \, (a-c+3) \nu + \eta}
	  +
	  2 e^{-\b b (J -J') + 2 \, (a-c+b+2) \nu + \eta }
	  \label{eq:5.22}
	 \\
	 &|\ve_{2}| \leq
	 |\ve'| +
	 2 e^{- \b( J- J')-2 \b b J' + 2\, (c+2) \nu +\eta}
	\label{eq:5.23}
\end{align}

Therefore for $\b$ large the sign of \eqref{eq:5.21}
will be given by the sign of the difference
\[
E^{(a-c+1)} - E(\d_{0})
=(a-c+1)(J-J')+2J -c(J-J')+2bJ'
\]
More precisely
\begin{align}
	%\st_{AW}(c+1)-\st_{AW}(c)
	\D\tau(c+1)-\D\tau(c)
	&\leq
	A  \bigg[ \frac{a+4}{2}-c -\frac{(b+1)K_{1}-J'}{J-J'}
	- \frac{\log \frac{|1-|\ve_{1}||}{1+|\ve_{2}|}}{2\b (J-J')}
	 \bigg] 
	\label{eq:5.24}
	\\
	=
	A  \bigg[ \frac{a+5}{2}&-c -\frac{(b-1)J'+2(b+1)K_{1}}{J-J'}
	- \frac{\log \frac{|1-|\ve_{1}||}{1+|\ve_{2}|}}{2\b (J-J')}
	\bigg] 
	\label{eq:5.25}
	\\
	%\st_{AW}(c+1)-\st_{AW}(c) 
	\D\tau(c+1)-\D\tau(c)
	&\geq
	B  \bigg[  \frac{a+4}{2}-c +\frac{J'-(b+1)K_{1}}{J-J'}
	- \frac{\log \frac{|1+|\ve_{1}||}{1-|\ve_{2}|}}{2\b (J-J')}
   \bigg]
	\label{eq:5.26}
	\\
	&=
	B \bigg[\frac{a+3}{2}-c +\frac{(b+1)J'}{J-J'}
	- \frac{\log \frac{1+|\ve_{1}|}{|1-|\ve_{2}||}}{2\b (J-J')}
	\bigg]
	\label{eq:5.27}
\end{align}
where
$A= 2 \b (J-J')(1+|\ve_{2}|) e^{-\b c(J-J')+2\b b J'}$
and
$B= 2 \b (J-J') (1+ |\ve_{1}|) e^{-\b (a-c+1)(J-J') - 2\b  J}$.
On the other hand 
when 
$2c \leq a $ if $a$ is even or
$2c \leq a +1$ if $a$ is odd, we get from 
\eqref{eq:5.14}--\eqref{eq:5.20}
\begin{multline}
		\D\tau(c+1)-\D\tau(c)
		\geq
		e^{-\b c(J-J')}
		\Big[ \, e^{2\b b J'}(1-|\ve'|) + e^{-2\b  J}\chi(2c< a+1)
		\\
		-2 e^{- \b( J- J') + 2\, (c+1) \nu +\eta}
		-6 e^{-  \b(3 J- J') + 2\, (c+2) \nu +\eta}\, \Big]
		\label{eq:5.28}
	\end{multline}
	
	Therefore,
	for $a$ odd, inequality \eqref{eq:5.25} proves Statement~c) 
	of the theorem while \eqref{eq:5.27}  gives 
	$\D \tau (\frac{a+3}{2})< \D \tau (\frac{a+5}{2})$
	i.e. Statement~b) for $c= (a+3)/2$ the result for the other value 
	following from \eqref{eq:5.28}, all this provided $\b $ is large enough
	as stated in the hypotheses of the theorem.
	When $a$ is even we can conclude only if 
	$J' \not= (b+1)K_{1}$.
	In this case, the statement~c) follows  from \eqref{eq:5.24}
	while the statement~b) follows from \eqref{eq:5.26} and \eqref{eq:5.28}.
	
	To find the lower bound on $\b$,
	we let $\ve_{1}(\b M)$ and $\ve_{2}(\b M)$ be respectively the 
	upper bounds on $|\ve_{1}|$ and $|\ve_{2}|$ obtained by replacing
	$J'$ by $M/(b+1)$ and $K_{1}$ by  $M/2(b+1)$ in \eqref{eq:5.22}
	 and \eqref{eq:5.23}. We use also
	 $K_{c}-\nu c\geq M/(b+1)-(a-1)\nu$.
	 Then, we take 
	 $\b M /(b+1) = (a+b) \nu (\alpha) + (1/2)(\eta(\alpha)+\alpha)$.
	 The value of $\alpha$ giving the lower bound stated in the theorem 
	 ensures that 
	 $(1/2 \b M)\{\log [1 + \ve_{1}(\b M)]-\log [1 - \ve_{2}(\b M)]\} < 1$,
	 $\ve_{1}(\b M)<1$ and $\ve_{1}(\b M)<1$.
	
	The above analysis leads  also to:
	\begin{align}
		&\D\tau(r,1) = r (J_{AW}-J_{BW}) - \frac{ e^{- \b( J- (1+2b)J')}}{\b a}
		+ O(e^{- \b( J- (2b-1)J')})
		\label{eq:5.29}
		\\
		&\D\tau(r,a-1) = r (J_{AW}-J_{BW}) - \frac{2 e^{- 2\b( J- J')}}{\b a}
		\, \chi(a\geq 3)
		+ O(e^{- 3\b( J- J')})
		\label{eq:5.30}
	\end{align}
	The second term in the R.H.S.\ of \eqref{eq:5.29} comes from the 
	energy of the excitation
	$\d_{0}$ for $c=1$ and the second term in the R.H.S.\ of 
	\eqref{eq:5.30} comes from the 
	energy of the excitation
	$\d^{(1)}(0,0)$.
	Relations \eqref{eq:5.29} and \eqref{eq:5.30}  give Statement~a) and end
	the proof of Theorem~\ref{thm.5.1}. \quad
	\qed
\end{proofof}

Notice that by  \eqref{eq:5.29} the first order term for the corrections 
of Wenzel's law is given
 in the case $c=1$ by 
 $ (-1 / \b  a )  \exp \{ \b[ a(r-1)( J_{AW}-J_{BW}) - J_{AB}+J_{AW}-J_{BW}] \}$
 and thus decreases with the roughness $r$.

\section{Concluding remarks}

Within the $1+1$--dimensional SOS model, we have shown that the shape of a sessile 
drop on a rough substrate is given by the Winterbottom's construction.
Using low temperature expansions we have analyzed the first order 
corrections to the Wenzel's law  describing the contact angle $\t$ of 
this sessile drop. These corrections vanish as the temperature goes to 
zero.
For a given family of two steps substrate $(0,b)$ with the same 
roughness, 
we have shown that there is an optimal geometry that maximize or 
minimize the wall tension.
These results confirm and explain previous numerical simulations.
That is to say that, at least for this given family of substrates, 
there is indeed an optimal geometry for wetting.
That this property can be proved for more realistic models or 
experimentally is an open challenging question.

\section*{Acknowledgements}

J.D.C.\ is thankful to the Centre de Physique Th\'eorique--CNRS for 
partial support.
S.M.S.\  and J.R.\  wish to thanks the Centre de Recherche en 
Mod\'elisation Mol\'eculaire--Universit\'e de Mons--Hainaut 
for warm hospitality and  financial
support.

\section*{Appendix}
\renewcommand{\theequation}{A.\arabic{equation}}
\setcounter{equation}{0}

\noindent {\bf Lemma}\quad
\it 
The partition function
$\pfI$
satisfies the subadditivity property
\begin{equation} 
\pfI ( 2N, 2(V' + V''), M' + M'' ) \ge 
\pfI ( N, V', M')\  Z_3 ( N, V'', M'') \  
e^{- 2 \beta \frac{|M''| }{2N-1} } 
\label{17} 
\end{equation}
\rm

\bigskip

\noindent 
{\it Proof.\/}
In order to prove this property we
associate a configuration ${\confh}$
of the first system in the box of length $2N$,
to a pair of configurations ${\confh}'$ and ${\confh}''$
of the system in  a box of length $N$, as follows
\begin{equation}
\begin{array}{rclr}
h_{2i}
&=&
\displaystyle h'_{i} + h''_{i} \; , 
& i=0,\dots , N
\\
h_{2i-1}
&=&
\displaystyle h'_{i-1 } + h''_{i} \; , 
& i=1,\dots , N 
\end{array}
\label{18}
\end{equation}
Then 
$h_{2N} = h'_{N} + h''_{N} = 0$,  
$h_{0} = h'_{0} + h''_{0} = M' + M''$ 
and
\beqn
 V(\confh)
&=&
2 \sum_{i=0}^N h'_i 
+ \sum_{i=0}^N h''_i 
+ \sum_{i=1}^N h''_i 
\nonumber
\\
& =& 2\  [ V({\confh'}) + V({\confh''}) ] - M'' 
\nonumber
\eeqn
This shows that the configuration $\confh$ belongs to
$ \pfI ( 2N, 2(V' + V'') - M'', M' + M'') $.  
Since $ H_{2N}({\confh}) = H_N ({\confh'}) + H_N ({\confh''})  $,
because 
$n_{2i} =n'_i$ 
and 
$n_{2i-1} =n''_i$ 
for $i=1,...,N-1 $, as follows from (\ref{18}),
we get 
$$
\pfI ( N, V', M')\  Z_3 ( N, V'', M'') 
\leq 
\pfI ( 2N, 2(V' + V'') - M'', M' + M'' )
$$
Then we use the change of variables
$\tilde{h}_i = h_i + [M''/(2N-1)]$
for $i=1,\dots,2N-1$,
$\tilde{h}_0 = h_0$, 
$\tilde{h}_{2N} = h_{2N}=0$ 
which gives
$$
\pfI ( 2N, V - M'', M )
\leq
e^{ 2 \beta |M''| / (2N-1) }
\ \pfI ( 2N, V , M )
$$
to conclude the proof. 
  \endproof

\end{document}